\documentclass[useAMS,usenatbib]{mn2e}
\usepackage{graphicx,amssymb}
\title[The origin and chemical evolution of carbon in the Galactic thin and 
thick disks]{The origin and chemical evolution of carbon in the Galactic thin 
and thick disks\thanks{Based on observations collected at the European Southern
Observatories (ESO) at La Silla in Chile, proposal no.~073D-0620(A)} }

\author[T. Bensby and S. Feltzing]{T. Bensby$^{1}$\thanks{E-mail: 
tbensby@umich.edu (TB); sofia@astro.lu.se (SF)} 
and
S. Feltzing$^{2}$\footnotemark[2]\\ $^{1}$Department of Astronomy, 
University of Michigan, 830 Dennison Building, Ann Arbor, MI 48109-1042, 
USA \\ 
$^{2}$Lund Observatory, Box 43, SE-221\,00 Lund, Sweden }

\begin{document}

\date{Accepted. Received; in original form}

\pagerange{\pageref{firstpage}--\pageref{lastpage}} \pubyear{0000}

\maketitle

\label{firstpage}

\begin{abstract}

In order to trace the origin and evolution of carbon in the Galactic
disk we have determined carbon abundances in 51 nearby F and G dwarf
stars. The sample is divided into two kinematically distinct subsamples
with 35 and 16 stars that are representative of
the Galactic thin and thick disks, respectively.  The analysis is
based on spectral synthesis of the forbidden [C\,{\sc i}] line at
872.7\,nm using spectra of very high resolution ($R\approx 220\,000$)
and high signal-to-noise ($S/N\gtrsim 300$) that were obtained with
the CES spectrograph on the ESO 3.6-m telescope on La Silla in
Chile. We find that [C/Fe] versus [Fe/H] trends for the thin and thick disks
are totally merged and  flat for sub-solar metallicities. The
thin disk that extends to higher metallicities than the thick disk,
 shows a shallow
decline in [C/Fe] from $\rm [Fe/H]\approx0$ and up to $\rm
[Fe/H]\approx+0.4$. The [C/O] versus [O/H] trends are well separated
between the two disks (due to differences in the oxygen abundances)
and bear a great resemblance to the [Fe/O] versus [O/H] trends. Our
interpretation of our abundance trends is that the sources that are
responsible for the carbon enrichment in the Galactic thin and thick
disks have operated on a time-scale very similar to those that are
responsible for the Fe and Y enrichment (i.e., SN\,Ia and AGB stars,
respectively).  We further note that there exist other observational
data in the literature that favour massive stars as the main sources
for carbon. In order to match our carbon trends, we believe that the
carbon yields from massive stars then must be very dependent on
metallicity for the C, Fe, and Y trends to be so finely tuned in the
two disk populations. 
Such metallicity dependent yields are no longer
supported by the new stellar models in the recent literature. 
For the Galaxy we hence conclude 
that the carbon enrichment at  metallicities typical 
of the disk is mainly due to low and intermediate mass stars, while 
massive stars are still the main carbon contributor at low metallicities 
(halo and metal-poor thick disk).
\end{abstract}


\section{Introduction}

Next to hydrogen, helium and oxygen, carbon is the most abundant
element in the Universe and plays an important r\^ole in the chemical
evolution of galaxies. Although the nuclear process (helium burning)
that is responsible for the formation of carbon atoms is well-known
\citep[see, e.g., reviews by][]{wallerstein1997, arnett2004}, it is
unclear which objects that contribute to the carbon enrichment of the
interstellar medium. This has been much debated during recent years;
\cite{gustafsson1999} conclude that carbon is mainly contributed from
massive, metal-rich stars, and not from low-mass stars;
\citet{chiappini2003, chiappini2003b} find strong indications that carbon 
is produced
in low and intermediate mass stars; \cite{shi2002} find that carbon
is contributed by super-winds from massive stars in the
early stages of disk formation in the Galaxy, while at later stages significant
amounts of carbon are contributed by low-mass stars. Other studies
that favour a mixture of low mass, intermediate mass, and high mass stars
include \cite{liang2001, gavilan2005, carigi2005} while
\cite{henry2000} and \cite{akerman2004} suggest massive stars to be
the main source.

Only few previous studies \citep{andersson1994, gustafsson1999,
takeda2005} have based their carbon abundances on the forbidden
[C\,{\sc i}] line at 872.7\,nm. This line is highly insensitive to errors
in the stellar model atmosphere \citep[see, e.g.,][]{asplund2005} and
gives carbon abundances of high accuracy in solar-type stars, even
when using one-dimensional (1-D) models under the assumption of local
thermodynamic equilibrium (LTE).  Unfortunately, the line is very weak
and is blended with a weak feature that is likely to be an Fe\,{\sc i}
line \citep[e.g.][]{lambert1967}.  So, in order to achieve accurate
abundances from this line it is desirable to have spectra of 
high-resolution and high signal-to-noise ($S/N$) and it is 
essentially only the \cite{gustafsson1999} study that has reliable
carbon trends based on the forbidden line (see discussion in 
Sect.~\ref{sec:systematicerrors}). They did, however, not
include the blending Fe\,{\sc i} line in their analysis.

As a step to achieve a deeper understanding of the sites of carbon
formation and the chemical histories of the Galactic disks we have
obtained high-resolution, high S/N spectra of the forbidden [C\,{\sc
i}] line at 872.7 nm for 51 dwarf stars in the solar
neighbourhood. These stars are already well studied, have well
determined oxygen and iron abundances, and have known kinematics
\citep{bensby2003, bensby2004, bensby2005}.  That the kinematics are
known is important as this makes it possible to properly investigate
the carbon trends in the two disks, independent of each other, and to
perform a differential analysis of the carbon trends in the thin and
thick disks in order to investigate the differences and similarities
in the disks' chemical histories. The latter investigation could,
combined with other evidence, be important for understanding the
time-scales involved in the chemical enrichment of the gas that today
make up what we observe as the stars in the Galactic thick disk.

In Sect.~\ref{sec:sample} we shortly describe the stellar sample and
the observations. Sect.~\ref{sec:reduction} discusses the data
reduction and associated problems and Sect.~\ref{sec:abundances} discusses
 the abundance analysis and estimate random and systematic
errors in the derived carbon abundances. In Sect.~\ref{sec:discussion}
we describe our resulting carbon trends and we use our results to
trace the possible sources of carbon as well as discuss the chemical
evolution of the Galactic disks. Finally, Sect.~\ref{sec:summary}
concludes with a summary of our findings.

\begin{figure}
\centering
\resizebox{0.9\hsize}{!}{\includegraphics[bb=100 0 370 80, clip]{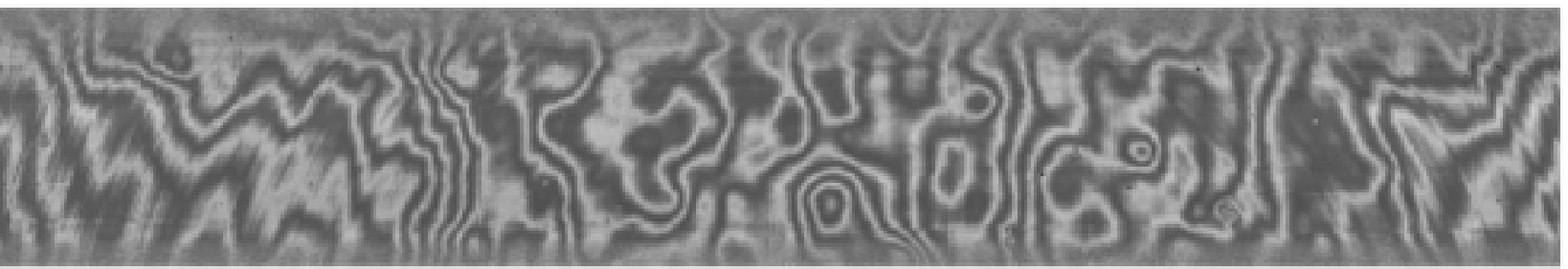}}
\resizebox{0.9\hsize}{!}{\includegraphics[bb=100 0 370 80, clip]{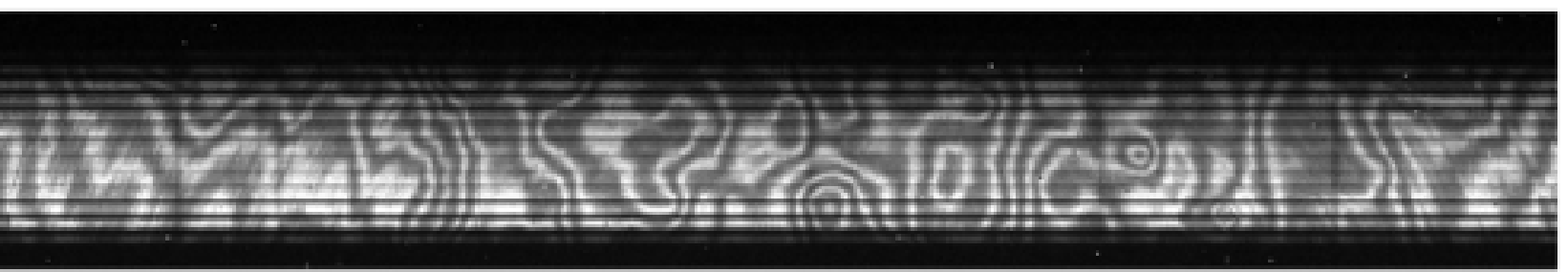}}
\resizebox{0.9\hsize}{!}{\includegraphics[bb=100 0 370 80, clip]{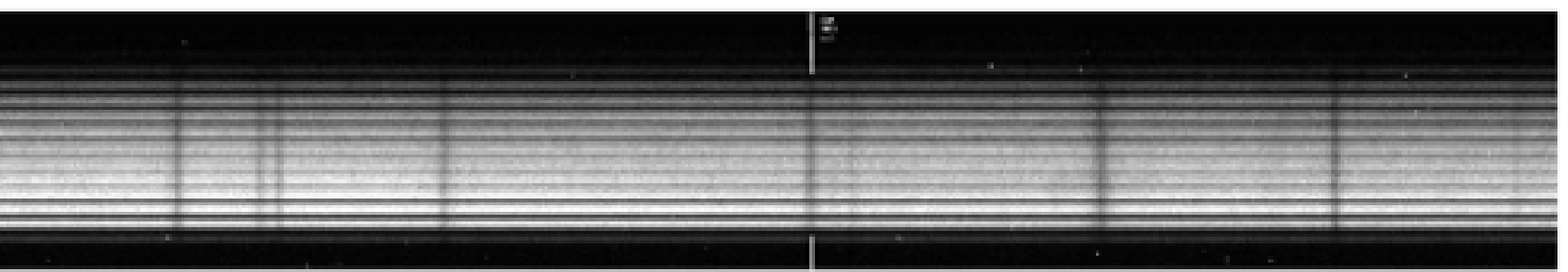}}
\caption{
        {\bf Top:} A normalised flatfield frame.
        {\bf Middle:} A ``raw" science frame (in this case a solar spectrum)
        {\bf Lower:} The flatfielded science frame. The strong Si\,{\sc i} line
        at 872.8\,nm, in whose left wing the [C\,{\sc i}] line is
        located, has been marked. The full CCD frames are 4096x1024 pixels.
        Here we show the central parts (approximately 2000x500 pixels
        in size) from $\sim 871.6$\,nm to $\sim 874.0$\,nm.
        }
\label{fig:ces}
\end{figure}
\begin{figure}
\resizebox{\hsize}{!}{\includegraphics{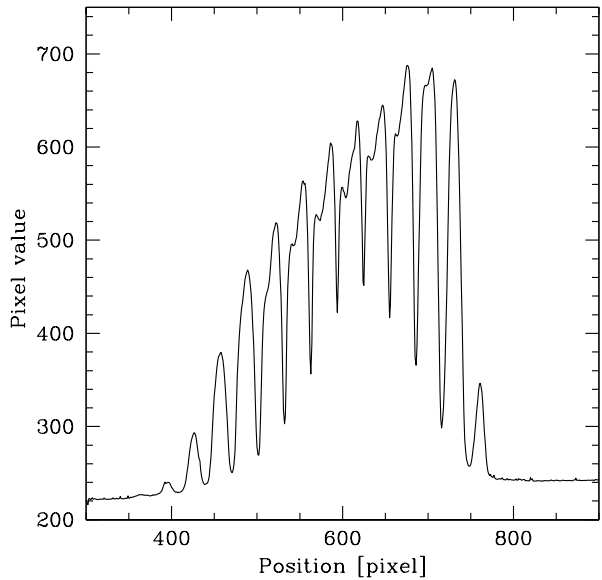}}
\caption{The average profile (based on 4096 rows) perpendicular
        to the dispersion direction. Note that the background
        level is higher on the right hand side (pixel value $\sim 240$)
        than on the left hand side (pixel value $\sim 220$).
        }
\label{fig:profile}
\end{figure}

\section{Stellar sample and observations} \label{sec:sample}

The stellar sample (listed in Table~\ref{tab:sampleA}) 
contains 51 F and G dwarf stars (35 thin disk and 16 thick disk stars) 
which form a subset of the samples in our previous studies
\citep[in total 102 stars, see][]{bensby2003, bensby2004, bensby2005}.
The selection criteria used to assign
a star to either the thin or the thick disk as well as further
details and discussions about the kinematic properties of the stars
in the sample can be found  in \cite{bensby2003, bensby2005} and in
Sect.~\ref{sec:kinematic}.

Observations were carried out by TB during six nights in September
2004 with the Coud\'e Echelle Spectrograph (CES) on the ESO 3.6-m
telescope on La Silla in Chile. All spectra were obtained with the
highest resolution $R\approx 220\,000$. Although our exposure times
were calculated to give spectra with signal-to-noise ratios that
exceed at least 350 this was unfortunately rarely the case in the
reduced spectra. The reason for this degradation in quality of the
reduced spectra is due to unexpectedly strong fringing patterns
from/in the CCD detector in this wavelength region.  The treatment of
this feature is discussed in more detail in Sect.~\ref{sec:reduction}.

Further, long exposures were split into shorter exposures of maximum
30 minutes. Solar spectra were obtained each day by  observing the
day-sky approximately one hour before sunset. Flatfield and bias
exposures were taken at the beginning and at the end of the nights and
additional flatfields at least once more during the nights. Wavelength
calibration spectra were obtained from a thorium-argon hollow cathode
lamp. The last four nights we also included exposures of rapidly
rotating B stars that were used in the reduction procedure to minimise
the residuals from the fringing pattern.

\begin{figure}
\resizebox{\hsize}{!}{\includegraphics{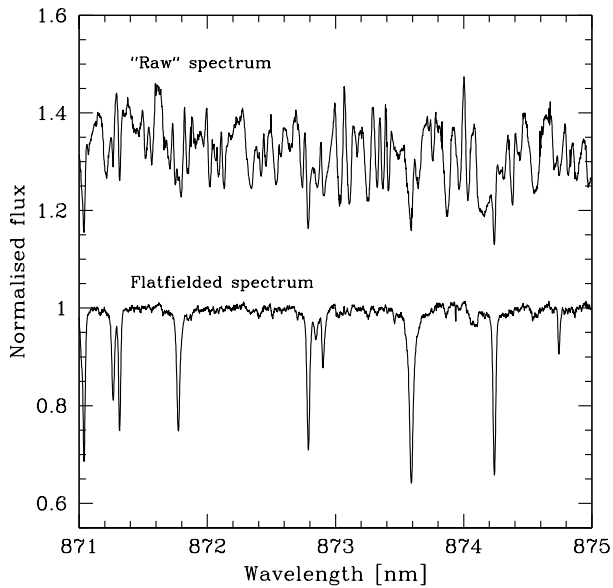}}
\caption{This comparison shows the efficiency of the flatfield in
        removing the fringing pattern from a ``raw" image.  The 1-D
        spectra shown here are the same as those in the two bottom
        panels in Fig.~\ref{fig:ces} with the exception that the
        reduced and flatfielded spectrum has been filtered for cosmic
        ray hits (which was not the case in  Fig.~\ref{fig:ces}). The
        ``raw" spectrum has also been normalised  and vertically
        shifted for this plot.  }
\label{fig:rawspectrum}
\end{figure}
\begin{figure}
\resizebox{\hsize}{!}{\includegraphics{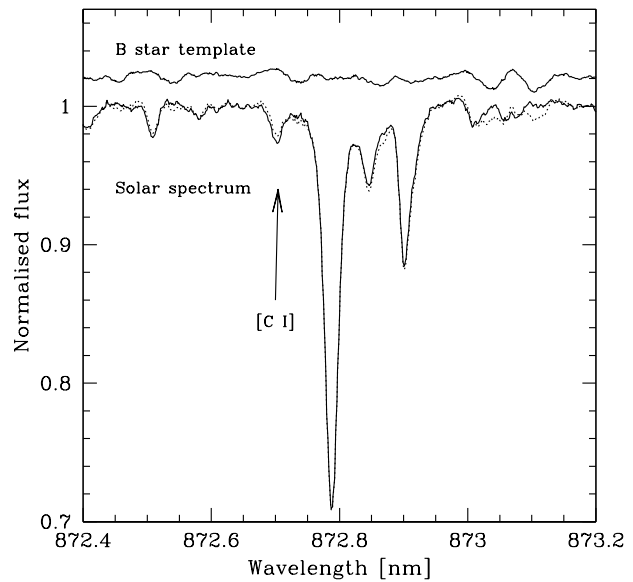}}
\caption{
        The top spectrum shows the B star template.
        It has been shifted vertically for this plot.
        Also shown is a reduced solar spectrum (dotted line,
        same spectrum as in Fig.~\ref{fig:rawspectrum})
        and the same solar spectrum but divided by the normalised
        B star template (solid line). 
        }
\label{fig:sun_bstar}
\end{figure}

\section{Data reduction}  \label{sec:reduction}

\begin{table*}
\centering
\caption{Stellar names, chemical abundances, and stellar
parameters. Abundances  are given relative to our solar
abundances. [C/H] abundances derived without the blending Fe\,{\sc i}
line are given in italics in parenthesis. Metallicities ([Fe/H]),
effective temperatures ($T_{\rm eff}$), surface gravities ($\log g$),
microturbulences ($\xi_{\rm t}$), and space velocities ($U_{\rm LSR}$,
 $V_{\rm LSR}$,  $W_{\rm LSR}$) are taken from Bensby et
al.~(2003, 2005).  Oxygen abundances ([O/H]) are taken from Bensby et
al.~(2004b) and are generally based on the forbidden [O\,{\sc i}] line
at 630.0\,nm unless otherwise indicated (``*" means that the permitted O\,{\sc i} 
triplet lines at
$\lambda\approx777$\,nm were used and ``**"  means that the
forbidden [O\,{\sc i}] line at $\lambda=636.3$\,nm was used). $v_{\rm
R-T}$ is the macroturbulent broadening (RAD-TAN) that we used (these
values have no ``real physical" meaning and are only listed for
completeness).}
\label{tab:sampleA} 
\setlength{\tabcolsep}{1.5mm}
\begin{tabular}{rrrrrrrccccrrr}
\hline
HIP & HD & [Fe/H] & [C/H] & {\it ([C/H])} & [O/H] & [C/O] & $T_{\rm eff}$ & $\log g$ & $\xi_{\rm t}$  & $v_{\rm R-T}$  & $U_{\rm LSR}$ &  $V_{\rm LSR}$ &  $W_{\rm LSR}$     \\
\noalign{\smallskip}
    &    &        &       &               &       &       &      [K]      &    [cgs] & [km\,s$^{-1}$] & [km\,s$^{-1}$] & [km\,s$^{-1}$] & [km\,s$^{-1}$] & [km\,s$^{-1}$]              \\
\hline
\multicolumn{13}{c}{\it THIN DISK STARS}                                                                                  \\
\noalign{\smallskip}
    910  &     693  &  $-0.36$  &  $-0.33$  &  {\it (--0.34)}  &  $     $  &  $     $  &  $6220$  &  $4.07$  &  $1.43$  & 5.09 & $  29.0 $  &  $   -7.8 $  &  $ -11.1$\\
   2235  &    2454  &  $-0.28$  &  $-0.11$  &  {\it (--0.13)}  &  $     $  &  $     $  &  $6645$  &  $4.17$  &  $1.75$  & 8.33 & $  23.2 $  &  $  -26.2 $  &  $  -6.7$\\
   2787  &    3229  &  $-0.11$  &  $-0.05$  &  {\it (--0.08)}  &  $     $  &  $     $  &  $6620$  &  $3.86$  &  $1.70$  & 5.44 & $ -16.5 $  &  $  -21.7 $  &  $ -12.9$\\
   3142  &    3735  &  $-0.45$  &  $-0.37$  &  {\it (--0.39)}  &  $-0.25$  &  $-0.12$  &  $6100$  &  $4.07$  &  $1.50$  & 4.13 & $ -35.5 $  &  $  -45.8 $  &  $  19.9$\\
   3909  &    4813  &  $-0.06$  &  $-0.02$  &  {\it (--0.03)}  &  $-0.11$  &  $ 0.09$  &  $6270$  &  $4.41$  &  $1.12$  & 3.85 & $  31.4 $  &  $    2.5 $  &  $  -4.8$\\
   5862  &    7570  &  $ 0.17$  &  $ 0.21$  &  {\it   (0.20)}  &  $ 0.06$  &  $ 0.15$  &  $6100$  &  $4.26$  &  $1.10$  & 4.96 & $ -34.1 $  &  $  -17.1 $  &  $  -6.0$\\
   7276  &    9562  &  $ 0.20$  &  $ 0.19$  &  {\it   (0.18)}  &  $ 0.14$  &  $ 0.05$  &  $5930$  &  $3.99$  &  $1.35$  & 4.38 & $   1.1 $  &  $  -21.1 $  &  $  19.5$\\
   9085  &   12042  &  $-0.31$  &  $-0.20$  &  {\it (--0.22)}  & *$-0.12$  &  $-0.08$  &  $6200$  &  $4.25$  &  $1.30$  & 5.43 & $ -42.5 $  &  $   -5.1 $  &  $   1.0$\\
  10306  &   13555  &  $-0.17$  &  $-0.03$  &  {\it (--0.05)}  &  $     $  &  $     $  &  $6560$  &  $4.04$  &  $1.75$  & 8.20 & $ -10.3 $  &  $   -6.8 $  &  $  11.2$\\
  12611  &   17006  &  $ 0.26$  &  $ 0.18$  &  {\it   (0.22)}  &**$ 0.14$  &  $ 0.04$  &  $5250$  &  $3.66$  &  $1.35$  & 3.47 & $  22.0 $  &  $  -13.5 $  &  $   0.1$\\
  12653  &   17051  &  $ 0.14$  &  $ 0.15$  &  {\it   (0.14)}  &  $ 0.04$  &  $ 0.11$  &  $6150$  &  $4.37$  &  $1.25$  & 6.67 & $ -21.3 $  &  $  -12.8 $  &  $  -3.1$\\
  14954  &   19994  &  $ 0.19$  &  $ 0.21$  &  {\it   (0.20)}  &  $ 0.11$  &  $ 0.10$  &  $6240$  &  $4.10$  &  $1.60$  & 9.24 & $  -9.7 $  &  $  -14.2 $  &  $   1.2$ \\
  15131  &   20407  &  $-0.52$  &  $-0.39$  &  {\it (--0.40)}  &  $-0.26$  &  $-0.13$  &  $5834$  &  $4.35$  &  $1.00$  & 2.97 & $   3.6 $  &  $   20.2 $  &  $  -9.6$\\
  17378  &   23249  &  $ 0.24$  &  $ 0.10$  &  {\it   (0.18)}  &  $ 0.13$  &  $-0.03$  &  $5020$  &  $3.73$  &  $0.80$  & 3.15 & $  -4.8 $  &  $   31.6 $  &  $  18.5$\\
  23941  &   33256  &  $-0.30$  &  $-0.20$  &  {\it (--0.22)}  & *$-0.23$  &  $ 0.03$  &  $6427$  &  $4.04$  &  $1.90$  & 10.0 & $  -2.0 $  &  $   -1.9 $  &  $   8.1$\\
  24829  &   35072  &  $ 0.06$  &  $ 0.09$  &  {\it   (0.07)}  &  $ 0.03$  &  $ 0.06$  &  $6360$  &  $3.93$  &  $1.70$  & 4.45 & $ -36.8 $  &  $  -26.3 $  &  $ -15.5$\\
  29271  &   43834  &  $ 0.10$  &  $ 0.08$  &  {\it   (0.11)}  &  $ 0.02$  &  $ 0.06$  &  $5550$  &  $4.38$  &  $0.80$  & 2.63 & $  29.5 $  &  $  -27.2 $  &  $  -5.8$\\
  78955  &  144585  &  $ 0.33$  &  $ 0.21$  &  {\it   (0.22)}  &  $ 0.17$  &  $ 0.04$  &  $5880$  &  $4.22$  &  $1.12$  & 3.16 & $ -16.6 $  &  $  -16.2 $  &  $  25.4$\\
  80337  &  147513  &  $ 0.03$  &  $-0.09$  &  {\it (--0.07)}  &  $ 0.03$  &  $-0.12$  &  $5880$  &  $4.49$  &  $1.10$  & 3.08 & $  24.0 $  &  $    4.0 $  &  $   5.8$\\
  80686  &  147584  &  $-0.06$  &  $-0.13$  &  {\it (--0.13)}  &  $-0.05$  &  $-0.08$  &  $6090$  &  $4.45$  &  $1.01$  & 3.99 & $  18.5 $  &  $   14.3 $  &  $   2.5$\\
  84551  &  156098  &  $ 0.12$  &  $ 0.07$  &  {\it   (0.05)}  &  $ 0.12$  &  $-0.05$  &  $6475$  &  $3.79$  &  $2.00$  & 7.61 & $ -31.9 $  &  $  -14.7 $  &  $  16.8$\\
  84636  &  156365  &  $ 0.23$  &  $ 0.21$  &  {\it   (0.21)}  &  $ 0.15$  &  $ 0.06$  &  $5820$  &  $3.91$  &  $1.30$  & 4.18 & $   1.7 $  &  $    3.8 $  &  $ -21.4$ \\
  86796  &  160691  &  $ 0.32$  &  $ 0.25$  &  {\it   (0.26)}  &  $ 0.14$  &  $ 0.11$  &  $5800$  &  $4.30$  &  $1.05$  & 3.44 & $  -5.0 $  &  $   -2.7 $  &  $   3.5$\\
  87523  &  162396  &  $-0.40$  &  $-0.33$  &  {\it (--0.35)}  &  $-0.31$  &  $-0.02$  &  $6070$  &  $4.07$  &  $1.36$  & 3.72 & $ -14.9 $  &  $   -5.2 $  &  $ -23.7$\\
  90485  &  169830  &  $ 0.12$  &  $ 0.18$  &  {\it   (0.17)}  &**$ 0.06$  &  $ 0.12$  &  $6339$  &  $4.05$  &  $1.55$  & 4.42 & $  -6.1 $  &  $    6.5 $  &  $  10.9$\\
  91438  &  172051  &  $-0.24$  &  $-0.30$  &  {\it (--0.28)}  &  $-0.18$  &  $-0.12$  &  $5580$  &  $4.42$  &  $0.55$  & 2.27 & $  47.1 $  &  $    3.0 $  &  $  2.9$\\
  99240  &  190248  &  $ 0.37$  &  $ 0.25$  &  {\it   (0.29)}  &  $ 0.19$  &  $ 0.24$  &  $5585$  &  $4.26$  &  $0.98$  & 3.08 & $ -38.2 $  &  $   -8.5 $  &  $  -8.2$ \\
 103682  &  199960  &  $ 0.27$  &  $ 0.25$  &  {\it   (0.25)}  &  $ 0.14$  &  $ 0.11$  &  $5940$  &  $4.26$  &  $1.25$  & 3.70 & $   3.8 $  &  $  -18.1 $  &  $   3.6$\\
 105858  &  203608  &  $-0.73$  &  $-0.60$  &  {\it (--0.62)}  &  $-0.46$  &  $-0.14$  &  $6067$  &  $4.27$  &  $1.17$  & 3.50 & $  -3.8 $  &  $   49.5 $  &  $  13.7$\\
 109378  &  210277  &  $ 0.22$  &  $ 0.10$  &  {\it   (0.15)}  &  $ 0.16$  &  $-0.06$  &  $5500$  &  $4.30$  &  $0.78$  & 2.60 & $  13.6 $  &  $  -45.5 $  &  $   1.8$\\
 110341  &  211976  &  $-0.17$  &  $-0.14$  &  {\it (--0.15)}  &  $-0.11$  &  $-0.03$  &  $6500$  &  $4.29$  &  $1.70$  & 6.64 & $   4.7 $  &  $   11.5 $  &  $   0.7$\\
 113137  &  216437  &  $ 0.22$  &  $ 0.22$  &  {\it   (0.22)}  &  $ 0.13$  &  $-0.09$  &  $5800$  &  $4.10$  &  $1.16$  & 3.99 & $  14.1 $  &  $   15.0 $  &  $   5.2$\\
 113357  &  217014  &  $ 0.20$  &  $ 0.13$  &  {\it   (0.15)}  &  $ 0.10$  &  $ 0.03$  &  $5789$  &  $4.34$  &  $1.00$  & 3.23 & $  -5.2 $  &  $  -23.2 $  &  $  22.0$\\
 113421  &  217107  &  $ 0.35$  &  $ 0.27$  &  {\it   (0.31)}  &  $ 0.16$  &  $ 0.11$  &  $5620$  &  $4.29$  &  $0.97$  & 2.75 & $   8.3 $  &  $   -3.8 $  &  $  18.3$\\
 117880  &  224022  &  $ 0.12$  &  $ 0.19$  &  {\it   (0.17)}  &  $ 0.05$  &  $ 0.14$  &  $6100$  &  $4.21$  &  $1.30$  & 5.36 & $ -36.2 $  &  $  -11.4 $  &  $   2.8$\\
\hline                                                                                                     
\multicolumn{13}{c}{\it THICK DISK STARS}                                                                                 \\
\noalign{\smallskip}                                                                       
   3086  &    3628  &  $-0.11$  &  $ 0.04$  &  {\it   (0.03)}  &  $ 0.09$  &  $-0.05$  &  $5840$  &  $4.15$  &  $1.15$  & 2.99 & $-159.4 $  &  $  -48.4 $  &  $  53.6$\\
   3497  &    4308  &  $-0.33$  &  $-0.25$  &  {\it (--0.25)}  &  $-0.05$  &  $-0.20$  &  $5636$  &  $4.30$  &  $0.80$  & 2.97 & $  60.0 $  &  $ -103.7 $  &  $ -19.0$\\
   5315  &    6734  &  $-0.42$  &  $-0.42$  &  {\it (--0.38)}  &  $ 0.00$  &  $-0.42$  &  $5030$  &  $3.46$  &  $0.86$  & 2.41 & $  60.3 $  &  $ -118.0 $  &  $  46.3$\\
  14086  &   18907  &  $-0.59$  &  $-0.53$  &  {\it (--0.52)}  &  $-0.16$  &  $-0.37$  &  $5110$  &  $3.51$  &  $0.87$  & 2.54 & $  18.6 $  &  $  -78.1 $  &  $ -13.1$ \\
  15510  &   20794  &  $-0.41$  &  $-0.38$  &  {\it (--0.35)}  &  $-0.01$  &  $-0.37$  &  $5480$  &  $4.43$  &  $0.75$  & 2.04 & $ -69.4 $  &  $  -89.3 $  &  $ -24.0$ \\
  17147  &   22879  &  $-0.84$  &  $-0.76$  &  {\it (--0.77)}  &  $-0.32$  &  $-0.44$  &  $5920$  &  $4.33$  &  $1.20$  & 2.29 & $ -99.0 $  &  $  -80.4 $  &  $ -37.0$\\
  18235  &   24616  &  $-0.71$  &  $-0.72$  &  {\it (--0.71)}  &  $-0.24$  &  $-0.48$  &  $5000$  &  $3.13$  &  $0.95$  & 2.97 & $ -17.0 $  &  $ -158.1 $  &  $ -20.1$ \\
  75181  &  136352  &  $-0.34$  &  $-0.20$  &  {\it (--0.21)}  &  $ 0.00$  &  $-0.20$  &  $5650$  &  $4.30$  &  $0.78$  & 2.58 & $-109.5 $  &  $  -41.4 $  &  $  43.3$ \\
  79137  &  145148  &  $ 0.30$  &  $ 0.12$  &  {\it   (0.24)}  &  $ 0.20$  &  $-0.08$  &  $4900$  &  $3.62$  &  $0.60$  & 3.03 & $  81.5 $  &  $  -49.8 $  &  $ -62.6$\\
  82588  &  152391  &  $-0.02$  &  $-0.11$  &  {\it (--0.02)}  &  $ 0.01$  &  $-0.12$  &  $5470$  &  $4.55$  &  $0.90$  & 3.32 & $  94.1 $  &  $ -105.5 $  &  $  16.6$  \\
  88622  &  165401  &  $-0.46$  &  $-0.44$  &  {\it (--0.44)}  &  $-0.05$  &  $-0.39$  &  $5720$  &  $4.35$  &  $0.80$  & 3.17 & $ -68.6 $  &  $  -84.7 $  &  $ -31.7$\\
  96124  &  183877  &  $-0.20$  &  $-0.07$  &  {\it (--0.06)}  &  $ 0.08$  &  $-0.15$  &  $5590$  &  $4.37$  &  $0.78$  & 2.27 & $ -28.0 $  &  $  -87.4 $  &  $ -15.4$\\
 103458  &  199288  &  $-0.65$  &  $-0.56$  &  {\it (--0.57)}  &  $-0.29$  &  $-0.27$  &  $5780$  &  $4.30$  &  $0.90$  & 2.60 & $  31.3 $  &  $  -96.3 $  &  $  52.9$\\
 108736  &  208998  &  $-0.38$  &  $-0.30$  &  {\it (--0.31)}  &  $-0.01$  &  $-0.29$  &  $5890$  &  $4.24$  &  $1.05$  & 3.00 & $ -15.6 $  &  $  -72.9 $  &  $  44.6$\\
 109821  &  210918  &  $-0.08$  &  $-0.08$  &  {\it (--0.08)}  &  $ 0.10$  &  $-0.18$  &  $5800$  &  $4.29$  &  $1.05$  & 2.72 & $ -36.9 $  &  $  -86.9 $  &  $  -2.5$\\
 118115  &  224383  &  $-0.01$  &  $ 0.07$  &  {\it   (0.07)}  &  $ 0.04$  &  $ 0.03$  &  $5800$  &  $4.30$  &  $1.00$  & 3.05 & $ -65.0 $  &  $  -77.9 $  &  $   3.3$\\
\hline
\end{tabular}
\end{table*}

The top panel in Fig.~\ref{fig:ces} shows a part of the CCD frame for
a (normalised) flatfield exposure  and the middle panel a solar
spectrum. As can be seen they are severely affected by
fringing. However, this fringing pattern was stable and did not change
over the six nights the observations lasted.  As the removal of this
pattern is extremely important for the quality of our results we give
a more detailed description of the reduction process.

The spectra were reduced using standard routines in the
MIDAS\footnote{ESO-MIDAS is the acronym for the European Southern
Observatory  Munich Image Data Analysis System which is developed and
maintained by the  European Southern Observatory.} software package
and consisted of the  following steps: \\
{\bf 1)} One master bias frame and several  (typically 4\,-\,5) master
flatfield  frames were constructed for each of the six nights by
averaging multiple exposures. Bias subtraction for the flatfields were
done after the bias value had been corrected for the difference
between its overscan value and that of the flatfield frame.  Since the
main method for removing fringing patterns is through the flatfielding
process care was taken when normalising the flatfield
frames. As proposed to be the best method in the CES
manual\footnote{Available at {\tt
http://www.ls.eso.org/lasilla/sciops/3p6/ces/}}  we normalised the
flatfield row by row, i.e., a low-order polynomial was fitted to each
row (i.e., along the dispersion direction) in the flatfield frame.
This process is necessary since the flatfields are obtained through
the fibre and the image slicer which results in a flatfield profile
that is composed of the same 12 slices as the science exposures (see
Fig.~\ref{fig:profile}). A plain division by this ``non-flat"
flatfield would result in too high weights given to the  low-flux
minima regions between the slices (i.e. dividing low numbers by  low
numbers). A normalised flatfield frame can be seen  in the top panel
in Fig.~\ref{fig:ces}.\\
{\bf 2)} The science frames were first corrected for the bias
level. Before dividing by a flatfield frame we also corrected the
frames for the fact that the background level is higher on the
right hand side than  on the left hand side of the profile perpendicular
to the dispersion direction (see Fig.~\ref{fig:profile}). This is due
to scattered light originating within the CES and do not contribute 
to the fringing pattern. If it is not
corrected for before the flatfielding the division by the flatfield
will overcompensate the pattern. We therefore fitted linear functions
between the left hand side (approximate pixel number 400) and the right
hand side (approximate pixel number 800) and made a subtraction so that the
two sides were level. This correction had a very positive effect on
the  quality of the reduced spectra for stars with long exposure times
as the  difference between the two sides relative to the total flux of
the object seem to increase with exposure time. For the flatfield
exposures the difference between left and right hand sides is practically
zero. After this correction we divided the
science frames with the normalised flatfield frames.  The middle panel
in Fig.~\ref{fig:ces} shows a ``raw" science frame and the bottom
panel in Fig.~\ref{fig:ces} an example of a frame that has been
corrected for the straylight pedestal and then divided by the
normalised flatfield frame shown in the top panel in
Fig.~\ref{fig:ces}.  In this way we were able to quite efficiently
remove much of the fringing pattern which is further illustrated in
Fig.~\ref{fig:rawspectrum}. The final spectra were also filtered for
cosmic ray hits (using the MIDAS task {\sc filter/cosmic}) before
dividing with the flatfield.\\
{\bf 3)} One-dimensional spectra were obtained by summing the fluxes
of all 12 slices (including the regions in between the slices, see
Fig.~\ref{fig:profile}).  The spectra were then wavelength calibrated
and re-binned to constant steps in wavelength.
Figure~\ref{fig:rawspectrum} shows the reduced solar spectrum.\\
{\bf 4)} We also observed several rapidly rotating B stars. Since
their spectral lines are all smeared out their spectra should more or
less form a featureless continuum. However, the reduced spectra for
our  19 B stars have a structure that show very little  variance when
compared to each other (apart from variations in the signal-to-noise
ratio).  This makes it most likely that this structure is the remains
of an ``imperfect" removal of the fringing and what we see in the B
star spectra should without any doubt also be present in all our other
reduced spectra.  In order to further reduce the residuals from the
fringing we therefore divided all spectra with a normalised B star
template. This B star template was derived as a weighted average of
all 19 B star spectra.  Figure~\ref{fig:sun_bstar} shows the
normalised B star template and the solar spectrum before and after
division with it.\\
{\bf 5)} For stars that have multiple exposures we then co-added the
individual spectra to one spectrum.  Finally, the spectra were
normalised by fitting a low-order Legendre polynomial to the continuum 
using the
IRAF\footnote{IRAF is distributed by  National Optical Astronomy
Observatories, operated by the Association  of Universities for
Research in Astronomy, Inc., under contract with the National Science
Foundation, USA.} task {\sc continuum}.

\section{Abundance determination} \label{sec:abundances}

\subsection{Stellar atmospheres}

We use one-dimensional, plane-parallel LTE  stellar model atmospheres
that were calculated with the Uppsala {\sc marcs} code
\citep{gustafsson1975, edvardsson1993, asplund1997}.   Surface
gravities ($\log g$) were determined from Hipparcos parallaxes,
metallicities ([Fe/H]) were determined using Fe\,{\sc i} lines,
effective temperatures ($T_{\rm eff}$) were determined by requiring
that the abundances derived from  Fe\,{\sc i} lines with different excitation
energies should all yield the same [Fe/H], 
and the microturbulence parameter ($\xi_{\rm t}$) by requiring
all Fe\,{\sc i} lines should yield the same 
abundances independent of line strength ($\log W_{\lambda}/\lambda$). 
All these parameters (listed in Table~\ref{tab:sampleA}) 
were determined in our previous studies \citep{bensby2003, bensby2005} 
wherein the iterative process to tune the stellar parameters 
also is fully described.

The [Fe/H] values listed in Table~\ref{tab:sampleA} are given relative to 
the solar Fe abundance that we derived using solar spectra obtained with
the same spectrographs as the stellar spectra 
\citep[see][]{bensby2003, bensby2005}. 
In these studies we also compare our derived [Fe/H] values as derived from
Fe\,{\sc i} lines to the [Fe/H] values derived from Fe\,{\sc ii} lines
(these we did not use in the tuning of the stellar parameters). 
Fe\,{\sc ii} lines are  more robust indicators than
Fe\,{\sc i} lines \citep[see][]{asplund2005araa} since they are
normally not affected by departures from LTE.
As we generally find good agreement between the [Fe/H] values derived from 
Fe\,{\sc i} and Fe\,{\sc ii} lines, respectively, 
(see Fig.~7 in \citealt{bensby2003}, and Fig.~2 in \citealt{bensby2005})
this indicates that for the range of parameters that the stars 
in our investigations
span the effects of departures from LTE are small, if at all measurable. 

\subsection{Synthetic spectra, atomic data, and line broadening} 
\label{sec:syntspec}

\begin{table}
\centering
\caption{Atomic data. The columns give: element and ionisation stage, 
         wavelength, lower
         excitation energy, $\log gf$-value, method for collisional
         broadening (see Sect.~\ref{sec:syntspec}), correction factor
         to the classical Uns\"old broadening, damping constants, and
         references for the  $\log gf$-values:  H93 = Hibbert et
         al.~(2003);  G97 = Galavis et al.~(1997); K93 = Kurucz~(1993); 
	BFL03 = Bensby et al.~(2003).  
	 (Note: see discussion in Sect.~4.4 and in 
	Allende Prieto et al.~(2002) regarding
	the Fe\,{\sc i} line and why $\log gf=-3.93$ is likely 
	an overestimate). The full line
         list that was used in the calculations of the synthetic
         spectra can be reproduced  by querying the VALD database ({\tt
         http://www.astro.uu.se/\~{}vald}) for lines in the region
         872.0\,--\,873.5\,nm.  }
\label{tab:atomdata}
\setlength{\tabcolsep}{1mm}
\begin{tabular}{ccrrcrcl}
\hline
        &  $\lambda$
        &  \multicolumn{1}{c}{$\chi_{\rm l}$}
        &  $\log gf$
        &  DMP
        &  $\delta \gamma_{6}$
        &  $\gamma_{\rm rad}$
        &  Ref.  \\
\noalign{\smallskip}
        &  (nm)
        &  \multicolumn{1}{c}{(eV)}
        &
        &
        &
        &  (s$^{-1}$)
        &        \\
\hline
[C\,{\sc i}] & 872.7126 & 1.264 & $-8.136$ & U & 2.50 & 1.0E+05 & H93, G97 \\
Fe\,{\sc i}  & 872.7132 & 4.186 & $-3.930$ & U & 1.40 & 6.1E+07 & K93      \\
Si\,{\sc i}  & 872.8010 & 6.181 & $-0.42*$ & C &      & 1.0E+05 & BFL03    \\
\hline
\end{tabular}
\end{table}

The synthetic spectra were calculated with the Uppsala {\sc spectrum}
software. As input it needs a stellar atmosphere model and a list of
spectral lines with atomic data.  To properly reproduce the region
around the [C\,{\sc i}] line we used the VALD database \citep{vald_3,
vald_2, vald_1} to extract all lines in the region 872.0\,nm to
873.5\,nm.  Table~\ref{tab:atomdata} lists the atomic data for the
[C\,{\sc i}] line at 872.7126\,nm \citep[$\log gf=-8.136$, theoretical
from][]{hibbert1993, galavis1997},  the blending Fe\,{\sc i} line at
872.7132\,nm \citep[$\log gf=-3.93$, theoretical from][]{kurucz1993},
and the Si\,{\sc i} line at 872.8010\,nm \citep[$\log gf=-0.42$,
astrophysical from][]{bensby2003}  whose left wing form the continuum
level for the [C\,{\sc i}] line.

The broadening of atomic lines by radiation damping was considered in
the determination of the abundances and the damping constants
($\gamma_{\rm rad}$) for the different lines were taken from the VALD
database.  Collisional broadening, or Van der Waals broadening, by
hydrogen atoms was  also considered. The width cross-section for the
Si\,{\sc i} line at 872.8010\,nm is taken from  \cite{barklem2}
(indicated by a ``C'' in Table~\ref{tab:atomdata}) and for the other
two lines we apply the correction term ($\delta \gamma_{6}$) to the
classical Uns\"old approximation for the Van der Waals damping
(indicated by a ``U'' in Table~\ref{tab:atomdata}).

\begin{figure}
\resizebox{\hsize}{!}{\includegraphics[bb=18 144 592 740, clip]{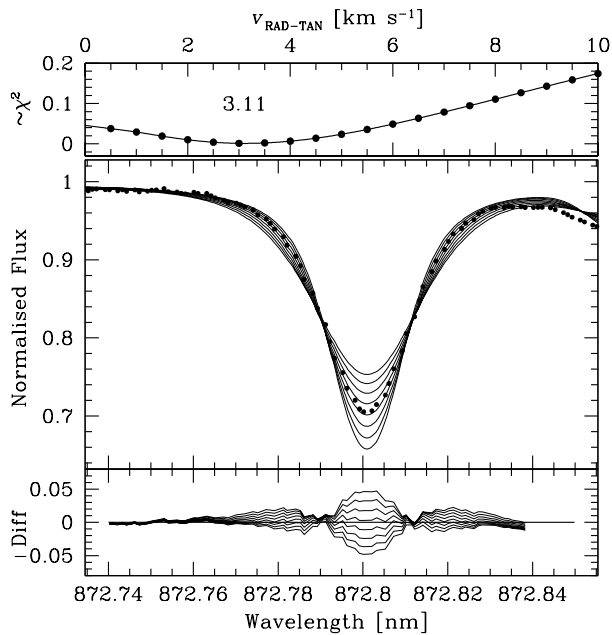}}
\caption{ Determination of the broadening due to macroturbulence. 
	(see Sect.~4.2). This plot shows the solar spectrum for
        which we  found a best matching RAD-TAN profile with a width
        of  3.11\,km\,s$^{-1}$.  The synthetic spectra shown in the
        two bottom panels have RAD-TAN widths from 1.5 to
        5.0\,km\,s$^{-1}$.  }
\label{fig:RADTAN}
\end{figure}

The synthetic spectrum is then convolved with a line profile to
reproduce the instrumental broadening and then with a another line
profile to reproduce the broadening due to  large-scale motions in the
stellar atmosphere.    The width of the instrument broadening profile
is set by the resolution of the spectra and was in our case
0.0042\,nm.  This value was determined by measuring the FWHMs in the
ThAr spectra and corresponds to a spectral resolving power of  about
$R\sim208\,000$ at 873.0\,nm ($R\equiv\lambda / \Delta\lambda = 873.0
/ 0.0042$).   The large-scale motions include macroturbulence  (that
has a radial-tangential, "RAD-TAN", profile) as well as the
line-of-sight component ($v\cdot\sin i$) of the stellar rotational
velocity.  The contribution from $v\cdot\sin i$  is usually small in
our types of stars (F and G dwarf stars)  in comparison to the
macroturbulence \citep[see, e.g.,][]{gray1992}.  To determine the
widths of the RAD-TAN profiles we used the quite strong Si\,{\sc i}
line at 872.8\,nm.  First we usually had to change the $\log gf$-value
of the Si\,{\sc i} line in order to get the line strength correct. We
then made 20 different synthetic spectra with different RAD-TAN
broadenings from 0 to 10 km\,s$^{-1}$.  The best fit was found by
minimising  an un-normalised $\chi^2$-function. As we here only want
to find the the best solution with a given set of free parameters we
need not normalise our $\chi^2$s.  An example of this is shown in
Fig.~\ref{fig:RADTAN}.

\begin{figure}
\resizebox{\hsize}{!}{\includegraphics{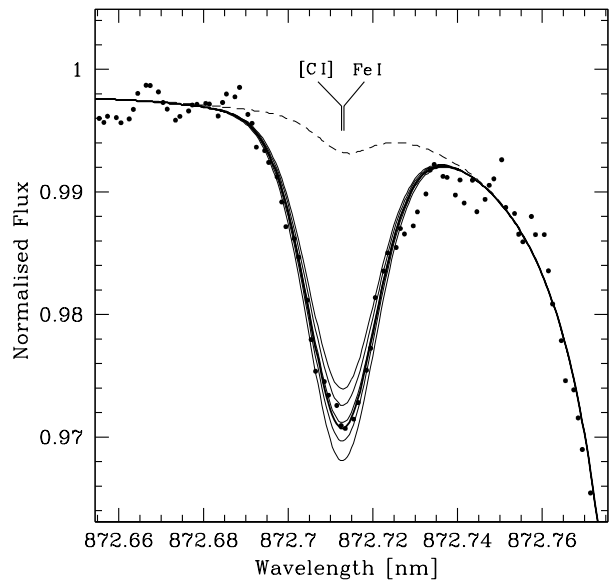}}
\caption{ The [C\,{\sc i}] line in the solar spectrum. The dots are
        the observed spectrum and the {\bf thick} 
	solid line the best
        fit representing a solar carbon abundance of  $\rm \log
        \epsilon_{\odot}(C) = 8.41$. The thin solid lines represents 5
        different carbon abundances from 8.34 to 8.46\,dex in steps of
        0.03\,dex. The dashed line shows the contribution from the
        blending Fe\,{\sc i} line. The wavelength positions of the
        [C\,{\sc i}] and the Fe\,{\sc i} lines are also indicated.  }
\label{fig:sun_c8727}
\end{figure}

\subsection{Carbon abundances}  \label{sec:carbonabundances}

\begin{figure*}
\resizebox{\hsize}{!}{
             \includegraphics[bb=18 144 592 730,clip]{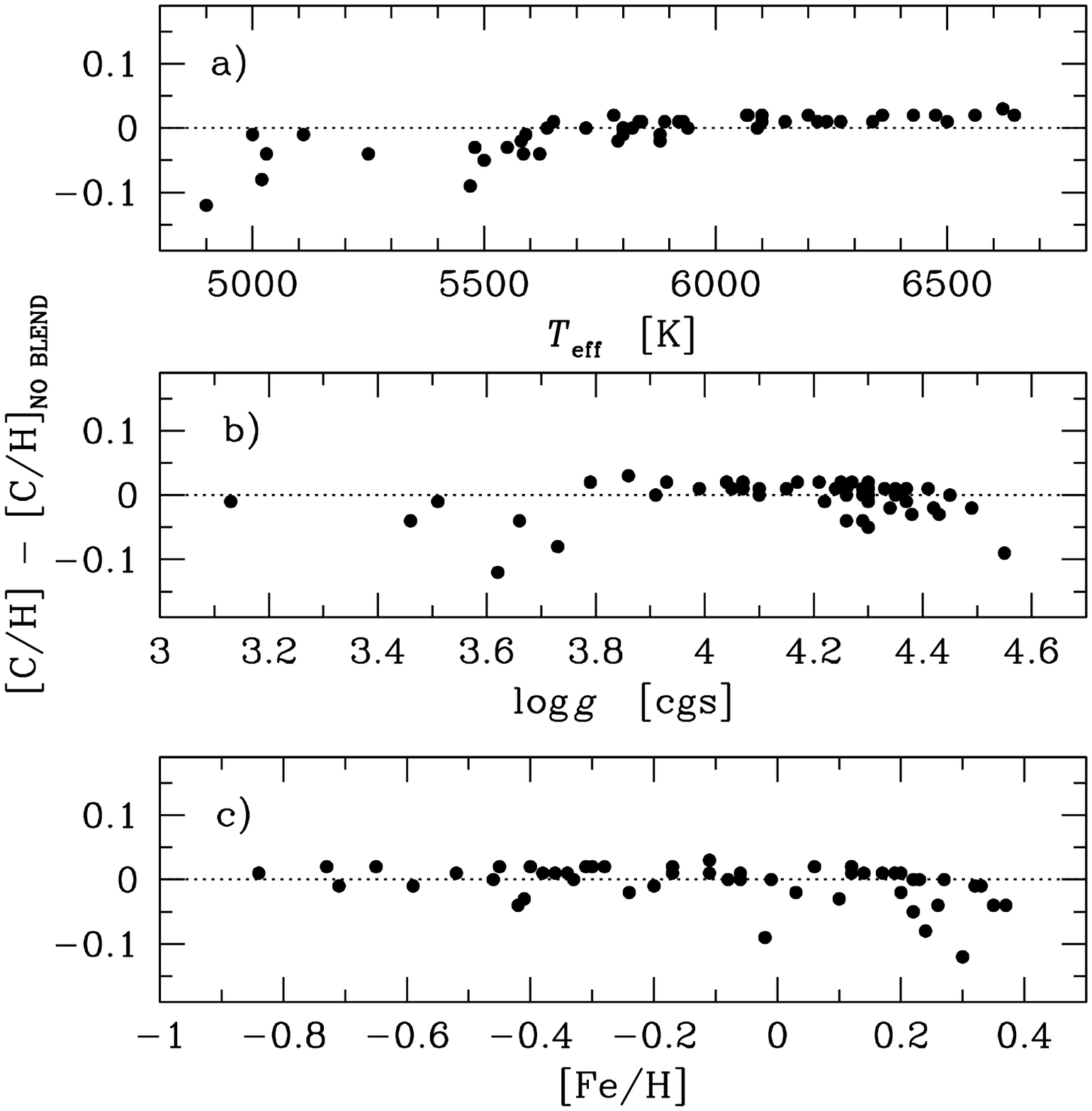}
             \includegraphics[bb=18 144 592 730,clip]{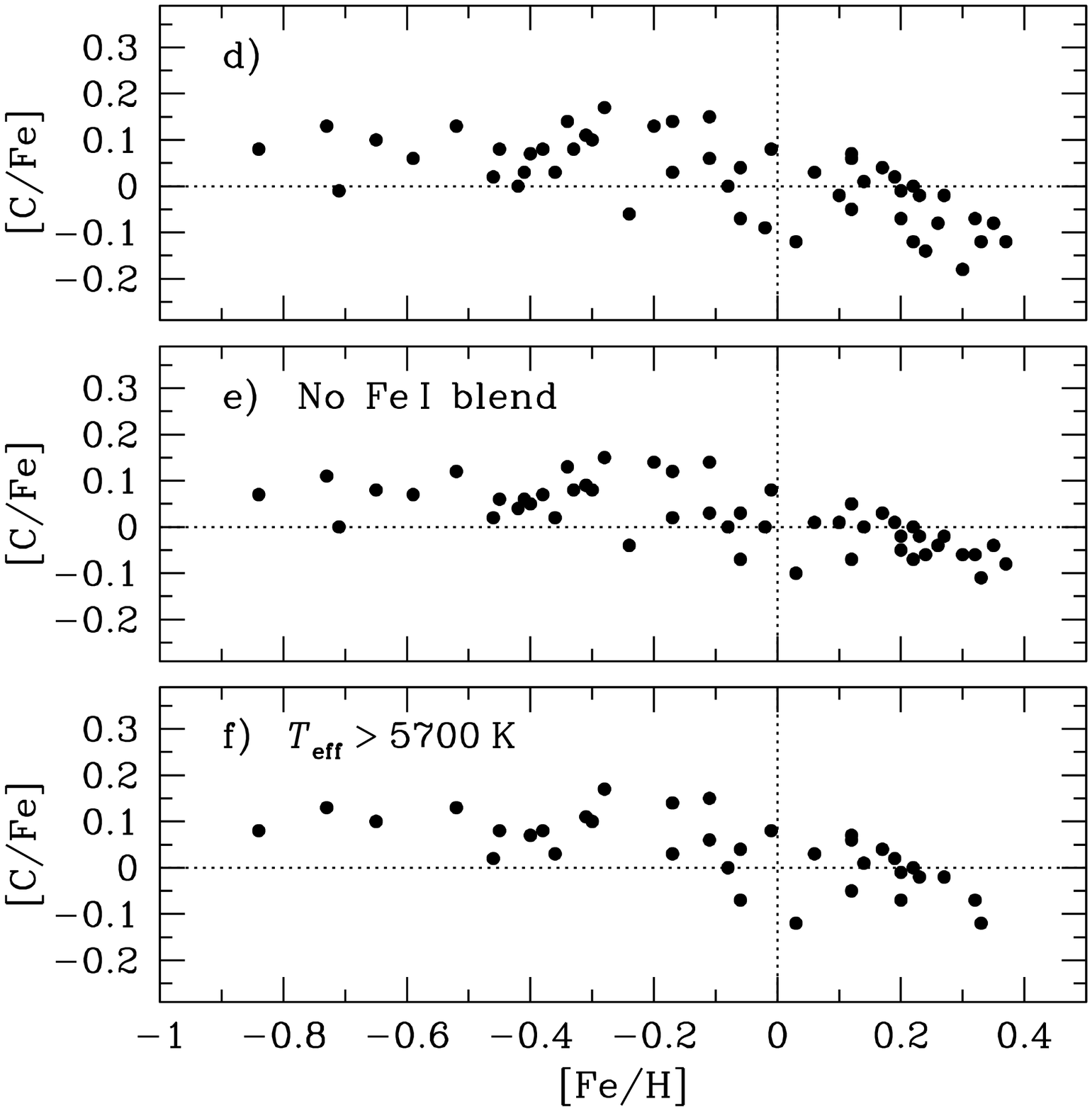}}
\caption{The effect on [C/H] when not including the Fe\,{\sc
         i} blend when synthesising the [C\,{\sc i}] line. The difference
        between the two derivations of carbon are shown in {\bf a)} versus
         effective temperature;  {\bf b)} versus surface gravity; and
         {\bf c)} versus metallicity. {\bf d)} then shows the
        [C/Fe] versus [Fe/H]
         trend for all stars; {\bf e)} [C/Fe] versus [Fe/H] for all
         stars when neglecting the Fe\,{\sc i} blend; and {\bf f)}
         [C/Fe] versus [Fe/H] (when including the Fe\,{\sc i} blend) 
         for stars that have $T_{\rm eff}>5700$\,K. }
\label{fig:chdiff_noblend}
\end{figure*}

The forbidden carbon line is located in the left wing of a Si\,{\sc i}
line and is blended by a line that most likely is an Fe\,{\sc i} line
\citep{lambert1967}.  The contribution from this blending line to the
joint [C\,{\sc i}]-Fe\,{\sc i} line profile is generally negligible at
sub-solar  metallicities while it in the Sun is estimated to be
between 0.1\,pm and 0.5\,pm \citep{allendeprieto2002}.  While most
previous studies have chosen to neglect this line in their abundance
analyses due to its assumed weakness and uncertain atomic data we have
included it as our sample contains many stars at super-solar [Fe/H]
where such a blend will grow in size. Since this Fe\,{\sc i} line only
has a calculated $\log gf$-value \citep{kurucz1993}, and could
therefore be associated with large errors, we must be careful and make
extensive checks. For comparison purposes we will therefore also
determine carbon abundances  neglecting the Fe\,{\sc i} blend.

By producing a set of synthetic spectra with carbon abundances
varying in steps of 0.03\,dex we then minimised an un-normalised
$\chi^2$-function to find the best fitting synthetic spectrum and
hence the carbon abundance (in the same way as we determined the
macroturbulence broadening, see Fig.~\ref{fig:RADTAN}). For seven
stars (HIP~10798, HIP~72673, HIP~81520, HIP~102264, HIP~107975,
HIP~3185, and HIP~109450) we did not determine carbon abundances due
to that their spectra showed defects probably as a result of an
insufficient removal  of the fringing pattern.

Figure~\ref{fig:sun_c8727} shows an example of the observed and the
synthetic spectra for the Sun. From our analysis of the solar spectrum
(using a MARCS model atmosphere with $T_{\rm eff}=5777$\,K, $\log
g=4.44$, $\xi_{\rm t} = 0.85$\,km\,s$^{-1}$, and RAD-TAN profile of
3.11\,km\,s$^{-1}$, see Fig~\ref{fig:RADTAN}) we get a solar carbon
abundance  of $\rm \log \epsilon_{\odot}(C) 8.41$  when the blending
Fe\,{\sc i} line is taken into account and $\rm \log
\epsilon_{\odot}(C) = 8.44$ when it is neglected.    
Both are in
good agreement with the recent analysis by \cite{allendeprieto2002}
and \cite{asplund2005} who found  a best value of $\rm \log
\epsilon_{\odot}(C) = 8.39$ using 3-D stellar atmosphere models and
$\rm \log \epsilon_{\odot}(C) = 8.41$ when using a MARCS model
atmosphere. These studies did, however, not include the blending
Fe\,{\sc i} line in their calculations.  The reason for that we see a
difference of 0.03\,dex (8.44 -8.41) between our and
\cite{allendeprieto2002} analysis, both using MARCS atmosphere models,
is probably due to uncertainties in the placement of the continuum
and/or differences between the spectra (that could originate from the
data  reduction process, see Sect.~\ref{sec:reduction}).  

Our final [C/H] abundances, both with and without the Fe\,{\sc i}
blend taken into account, are listed in
Table~\ref{tab:sampleA} (given relative to our solar abundance 
$\rm \log \epsilon_{\odot}(C) = 8.41$ and $\rm\log \epsilon_{\odot}(C)= 8.44$, 
respectively).

\subsection{Fe\,{\sc i} blend or no Fe\,{\sc i} blend?}

How does the inclusion of the Fe\,{\sc i} blend influence our final
carbon abundances and carbon trends? In
Figs.~\ref{fig:chdiff_noblend}a-c we show the difference in [C/H]
versus $T_{\rm eff}$, $\log g$, and [Fe/H], respectively. There is a
clear trend with $T_{\rm eff}$; the lower the temperature the larger
is the effect on [C/H] when neglecting the blend.  That the difference
in [C/H] generally is negative below 5700\,K and positive at higher
temperatures is due to that our abundance analysis is 
differential to the Sun.  At effective temperatures higher than $\sim
5700$\,K the difference in [C/H] is always smaller than   $\sim
0.03$\,dex no matter if the stars have high or low metallicities.  No
absolute trend with [Fe/H] can be seen (there is still a number of
stars  at high metallicities where the difference is negligible). 
From this we conclude that it is for a combination of 
low $T_{\rm eff}$ (and often low $\log g$), and high
[Fe/H] that we need to be extra careful and  correctly include the
Fe\,{\sc i} blend.

In Figs.~\ref{fig:chdiff_noblend}d-e we show the effect of the Fe\,{\sc
i}  blend on the final [C/Fe] versus [Fe/H] trend. Apart from a
somewhat lower spread in [C/Fe] when neglecting the blend the trends
are essentially  identical: both have a flat [C/Fe] trend at  $\rm
[Fe/H]<0$ and at $\rm [Fe/H]>0$ the [C/Fe] ratio declines when  going
to higher metallicities. Figure~~\ref{fig:chdiff_noblend}f shows the
trend (derived {\it with} the blend) but for stars that have $T_{\rm
eff}>5700$\,K and for which the effect on [C/Fe] due to an erroneous
treatment of the blend should be minimal (see
Fig.~\ref{fig:chdiff_noblend}a).  The trend seen in
Figs.~\ref{fig:chdiff_noblend}d-e still persist.

Summarising, we conclude that it is important to include the Fe\,{\sc
i} blend for stars that have low effective temperatures (below
5700\,K) and then especially at high [Fe/H]. It is, however, important
to remember that the $\log gf$-value that is available for the
Fe\,{\sc i} line is a theoretical one \citep{kurucz1993} and it is
possible that the calculated line strength for the blend is either
over- or underestimated.   It would therefore be very valuable to have
an accurate  experimental $\log gf$-value for this Fe\,{\sc i} line so
that it is possible to properly calculate its contribution to the
joint [C\,{\sc i}]-Fe\,{\sc i} line profile.  
We note that \cite{lambert1977} tried to estimate the $\log gf$-value
for this line by ensuring that the same C abundance is derived from 
the [C\,{\sc i}] line as from C$_{2}$ and CH lines.
They arrived at a value of $\log gf \approx -3.6$
which would lead to a higher Fe\,{\sc i} contribution to the joint
[C\,{\sc i}]\,-\,Fe\,{\sc i} line profile at 872.7\,nm than indicated
by $\log gf=-3.93$ from \cite{kurucz1993}. 
By looking at other Fe\,{\sc i} lines belonging
to the same multiplet \cite{allendeprieto2002} concluded in 
their analysis of the
solar spectrum that the contribution should be significantly
lower and hence that $\log gf = -3.93$ would overestimate
the line strength of the blending Fe\,{\sc i} line. 
Allende Prieto et al.'s~(2002) conclusion means that our carbon 
abundances might be somewhat
underestimated at high [Fe/H] since we used $\log gf=-3.93$
from \cite{kurucz1993}.
However, for our stellar sample we see that the general appearance 
of the [C/Fe] versus [Fe/H] trend is not affected by the blend (apart from a
somewhat smaller spread in [C/Fe] when {\it neglecting} it).
In what follows we therefore
continue to use our carbon abundances that we derived {\it with} the
Fe\,{\sc i} blend and these are the values we hereafter refer to
unless otherwise indicated.

\begin{figure}
\resizebox{\hsize}{!}{\includegraphics[bb=18 144 592 695,clip]{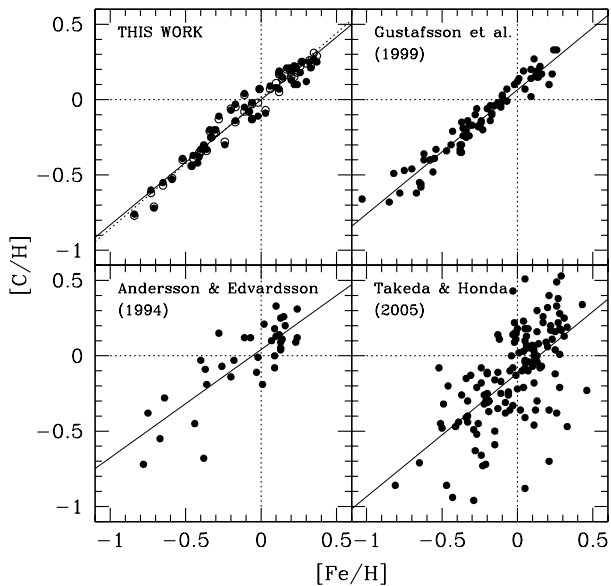}}
\caption{
        [C/H] versus [Fe/H] for our and the stellar samples from
        Gustafsson et al.~(1999), Andersson \& Edvardsson~(1994), and
        Takeda \& Honda~(2005). Linear regression lines are also plotted,
        see Eqs.~(\ref{eq:our})-(\ref{eq:th05}).
        For our stellar sample we also plot the abundances that were derived
        when neglecting the Fe\,{\sc i} blend (open circles and dotted line).
        }
\label{fig:ch_feh}
\end{figure}

\begin{figure*}
\resizebox{\hsize}{!}{
        \includegraphics[bb=18 144 592 725,clip]{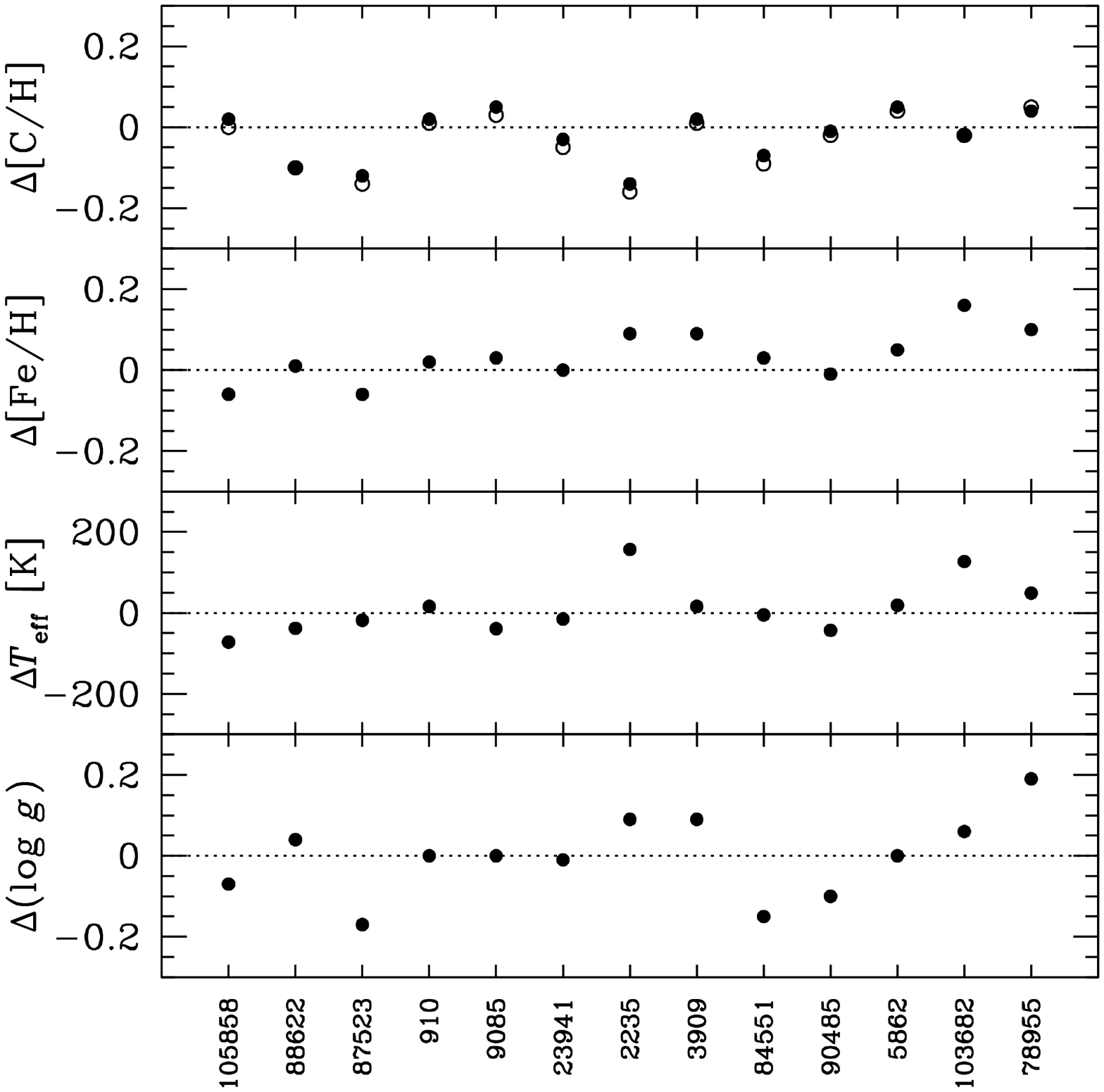}
        \includegraphics[bb=18 144 592 725,clip]{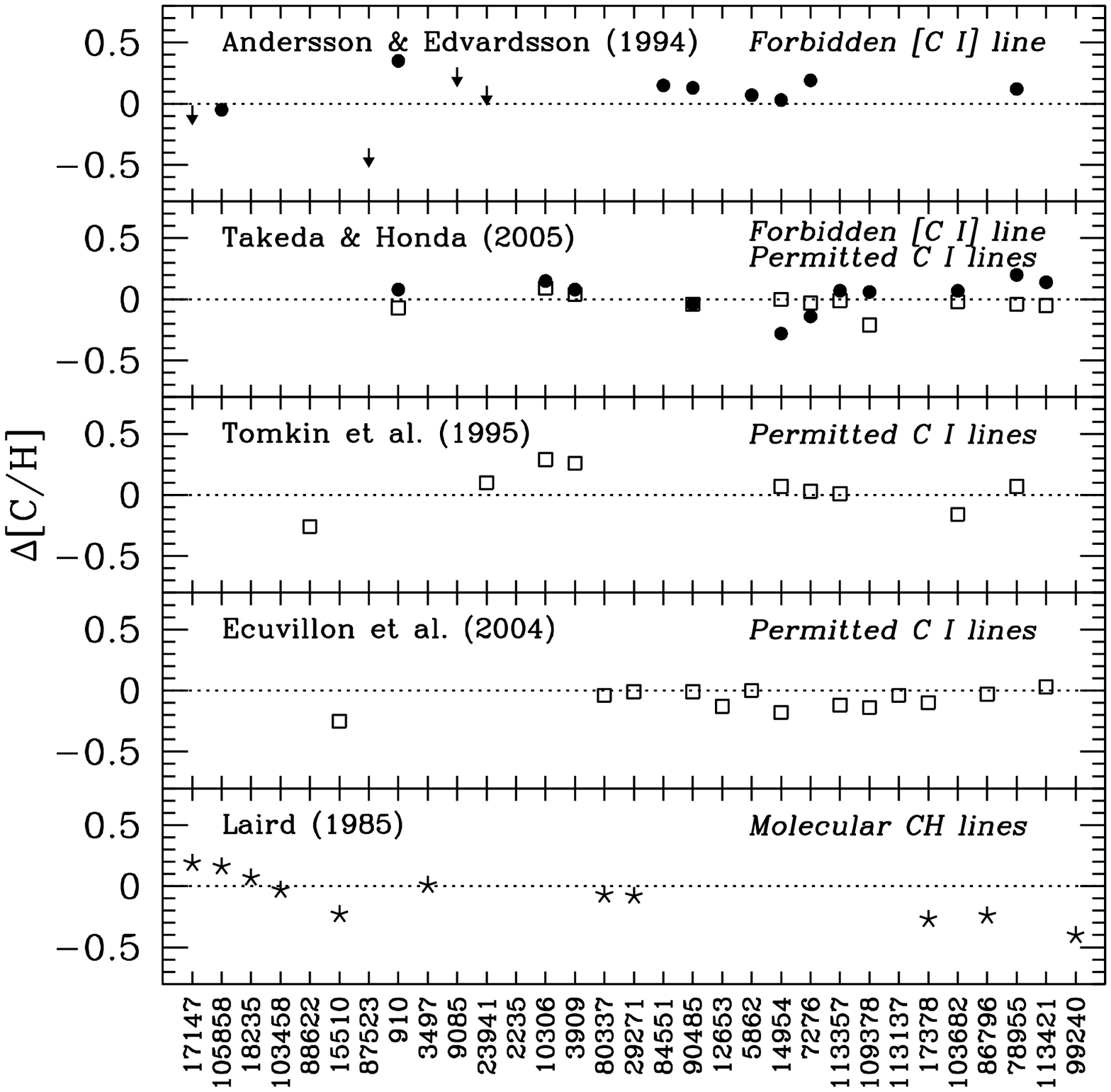}}
\caption{ {\bf Left hand panel:} A detailed comparison of abundances and
        stellar parameters for stars that we have in common with
        Gustafsson et al.~(1999).  For the $\rm \Delta [C/H]$ comparison we
        also include our abundances derived when neglecting the Fe\,{\sc i}
        blend (open circles). On the ``x-axis" we have marked the
        Hipparcos numbers of the stars. The stars have been sorted accorded
        to their metallicities, [Fe/H] increasing from the left to the right. 
        {\bf Right hand panel:} Comparison
        of carbon abundances for stars that we have in common with
        other studies. The figure shows our [C/H] values minus [C/H]
        values from the other studies and comparisons are made both
        with studies that use the forbidden [C\,{\sc i}] line as well
        as studies that use the permitted C\,{\sc i} lines and the
        molecular CH bands (as indicated in the figure). In the case
        of Andersson \& Edvardsson~(1994) the arrows mark
        stars that they only could give an upper estimate on [C/H]. In
        the case of Takeda \& Honda~(2005) the different symbols mark
        [C/H] abundances from the forbidden [C\,{\sc i}] line (solid
        circles) and from the permitted C\,{\sc i} lines (open
        squares).  The stars have been sorted accorded
        to their metallicities, [Fe/H] increasing to the right.
        On the ``x-axis" we have marked the Hipparcos
        numbers of the stars. Note the different scales on the "y-axes"
        in the left and right hand plots. The abundances from the different
studies have not been adjusted for differences in the stellar parameters.  }
\label{fig:carbon_comparison}
\end{figure*}

\subsection{Error analysis} 
\label{sec:systematicerrors}

To check for systematic effects we compare our derived carbon
abundances to other studies that also use the forbidden  [C\,{\sc i}]
line as  indicator of the carbon abundance 
\citep{gustafsson1999, andersson1994, takeda2005}. 
In Fig.~\ref{fig:ch_feh} we plot [C/H] versus [Fe/H] for
our data set as well as for data from these three studies. For each data 
set we have done a simple linear regression between [C/H] and [Fe/H].
These are  over-plotted in Fig.~\ref{fig:ch_feh}.  The regression lines
are given by the following equations (where G99$=$\citealt{gustafsson1999}; 
AE94$=$\citealt{andersson1994}; TH05$=$\citealt{takeda2005}):
\begin{equation}
\rm
[C/H] = [Fe/H] \cdot 0.83 - 0.00\hspace{10mm}(This\,work), \label{eq:our}
\end{equation}
\begin{equation}
\rm
[C/H] = [Fe/H] \cdot 0.87 + 0.01\hspace{10mm}(This\,work, \,no\,blend), \label{eq:ournoblend}
\end{equation}
\begin{equation}
\rm
[C/H] = [Fe/H] \cdot 0.82 + 0.07\hspace{10mm}({\rm G99}), \label{eq:g99}
\end{equation}
\begin{equation}
\rm
[C/H] = [Fe/H] \cdot 0.72 + 0.04\hspace{10mm}({\rm AE94}), \label{eq:a94}
\end{equation}
\begin{equation}
\rm
[C/H] = [Fe/H] \cdot 0.82 - 0.12\hspace{10mm}({\rm TH05}). \label{eq:th05}
\end{equation}

The tight relation between [C/H] and [Fe/H] that we see for our
stellar sample (see Fig.~\ref{fig:ch_feh}) is also present in the
\cite{gustafsson1999} data set (see also their Fig.~3).  As can be
seen the slopes ($\rm \Delta[C/H] / \Delta[Fe/H]$) are essentially the
same for our stellar sample and the \cite{gustafsson1999} and
\cite{takeda2005} samples. There are, however, offsets present between
these two data sets and ours. Relative to our relationship, the
\cite{gustafsson1999} relationship is shifted 0.07\,dex upwards (to
higher [C/H] values) while the \cite{takeda2005} relationship is
shifted 0.12\,dex downwards (to lower [C/H] values). The relationship
from the \cite{andersson1994} data set{\footnote{Many of the [C/H]
values in the \cite{andersson1994} data set are only upper limit
estimates. These were rejected from the analysis presented here.} is
shifted upwards by 0.04\,dex relative to our relationship and differs
from the other three in that it shows a shallower slope. The
\cite{andersson1994} and \cite{takeda2005} data sets have a
considerably larger spread. In the \cite{andersson1994} case the large
scatter is due to that their spectra was severely affected by fringing
and that many of their spectra had an insufficient $S/N$ to enable
good abundance determination from the extremely weak [C\,{\sc i}]
line. The reason for the large scatter in the \cite{takeda2005} data
is due to that their spectra in many cases were of low quality 
(Honda, private comm.) and in many weak-line cases they decided 
to adopt  the values from the permitted 
C \,{\sc i}  lines (average of $\lambda\lambda$505.2/538.0\,nm lines) as 
their final carbon abundance instead (note: Fig.~\ref{fig:ch_feh} and Eq.~\ref{eq:th05}
only include data from the forbidden line).

\begin{table}
\centering
\caption{Estimates of the effects on the derived abundances due to
         errors in the atmosphere parameters. $\sigma_{\rm cont}$ is the
	estimated error due to erroneous placement of the continuum
	(see text) and
        $\sigma_{\rm tot}$ is the total random error that were calculated
        assuming the individual errors to be uncorrelated. The  values
        for O and Fe were taken from Bensby et al.~(2004b).}
\label{tab:errors}
\centering
\setlength{\tabcolsep}{0.8mm}
\begin{tabular}{lrrrrrr}
\hline 
      & $\rm \left[\frac{Fe}{H}\right]$
      & $\rm \left[\frac{O}{H}\right]$
      & $\rm \left[\frac{O}{Fe}\right]$
      & $\rm \left[\frac{C}{H}\right]$
      & $\rm \left[\frac{C}{Fe}\right]$
      & $\rm \left[\frac{C}{O}\right]$ \\
 \hline
            \multicolumn{7}{c}{\bf HIP 88622, (--0.46, 5720, 4.35)} \\
 \noalign{\smallskip}
 $\Delta T_{\rm eff} = +70$ K                 & $+0.06$ & $+0.02$ & $-0.04$ & $-0.01$ & $-0.07$ & $-0.03$  \\
 $\Delta \log g = +0.1$                       & $-0.01$ & $+0.04$ & $+0.05$ & $+0.03$ & $+0.04$ & $-0.01$  \\
 $\Delta \xi_{\rm t} = +0.15$ km~s$^{-1}$     & $-0.02$ & $ 0.00$ & $+0.02$ & $ 0.00$ & $+0.02$ & $ 0.00$  \\
 $\rm \Delta [Fe/H] = +0.1$                   & $+0.01$ & $+0.03$ & $+0.02$ & $+0.01$ & $ 0.00$ & $-0.02$  \\
 \noalign{\smallskip}
 $\sigma_{\rm atm}$                           &   0.06  &   0.05  &   0.07  &   0.03  &   0.08  &   0.04   \\
\hline
            \multicolumn{7}{c}{\bf HIP 82588, (--0.02, 5470, 4.55)} \\
\noalign{\smallskip}
 $\Delta T_{\rm eff} = +70$ K                 & $+0.05$ & $+0.01$ & $-0.04$ & $-0.01$ & $-0.06$ & $-0.02$  \\
 $\Delta \log g = +0.1$                       & $-0.01$ & $+0.04$ & $+0.05$ & $+0.03$ & $+0.04$ & $-0.01$  \\
 $\Delta \xi_{\rm t} = +0.15$ km~s$^{-1}$     & $-0.03$ & $ 0.00$ & $+0.03$ & $ 0.00$ & $+0.03$ & $ 0.00$  \\
 $\rm \Delta [Fe/H] = +0.1$                   & $+0.01$ & $+0.04$ & $+0.03$ & $+0.02$ & $+0.01$ & $-0.02$  \\
\noalign{\smallskip}
 $\sigma_{\rm atm}$                           &   0.06  &   0.06  &   0.08  &   0.04  &   0.08  &   0.03   \\
 \hline

            \multicolumn{7}{c}{\bf HIP 103682, (+0.27, 5940, 4.26)} \\
 \noalign{\smallskip}
 $\Delta T_{\rm eff} = +70$ K                 & $+0.05$ & $+0.02$ & $-0.03$ & $ 0.00$ & $-0.05$ & $-0.02$  \\
 $\Delta \log g = +0.1$                       & $-0.01$ & $+0.05$ & $+0.06$ & $+0.04$ & $+0.05$ & $-0.01$  \\
 $\Delta \xi_{\rm t} = +0.15$ km~s$^{-1}$     & $-0.05$ & $ 0.00$ & $+0.05$ & $ 0.00$ & $+0.05$ & $ 0.00$  \\
 $\rm \Delta [Fe/H] = +0.1$                   & $ 0.00$ & $+0.03$ & $+0.03$ & $+0.01$ & $+0.01$ & $-0.02$  \\
 \noalign{\smallskip}
 $\sigma_{\rm atm}$                           &   0.07  &   0.06  &   0.09  &   0.04  &   0.09  &   0.03   \\
 \hline
 $\langle \sigma_{\rm atm}\rangle$            & $ 0.06$ & $ 0.06$ & $ 0.08$ & $ 0.04$ & $ 0.08$ & $ 0.03$  \\
 \hline
 $\sigma_{\rm cont}$                          &         &   0.03  &   0.03  &   0.06  &   0.06  &   0.07   \\
 \hline
 $\langle \sigma_{\rm tot}\rangle$            &   0.06  &   0.07  &   0.09  &   0.07  &   0.10  &   0.07   \\
 \hline
 \end{tabular}
\end{table}

It thus appear that it is safe to say that there is a systematic shift between
our abundances and those by \cite{gustafsson1999} since both data sets
are individually well defined. Whether this shift is due to the carbon
abundances (vertical shift in Fig.~\ref{fig:ch_feh}) or to the Fe
abundances (horizontal shift in Fig.~\ref{fig:ch_feh}), a combination
of both, or due to that we included the Fe\,{\sc i} blend and they did
not, will now be investigated by a comparison of individual abundances for
stars that we have in common.

The left hand panel in Fig.~\ref{fig:carbon_comparison} shows a detailed 
comparison for 13 stars that we have in common with 
\cite{gustafsson1999}. The mean difference in abundances and stellar 
parameters are (our values minus theirs, the uncertainties represent 
the one-sigma standard deviations):
$\rm    \Delta [C/H]   =  -0.02\pm0.07$\,dex,
$\rm    \Delta [C/H]_{NO\,BLEND}   =  -0.03\pm0.07$\,dex,
$\rm    \Delta [Fe/H]  =  +0.03\pm0.06$\,dex, 
$\rm    \Delta T_{\rm eff}  =  +12\pm66$\,K, 
$\rm    \Delta \log g  =  -0.00\pm0.10$\,dex.
Overall the mean differences between our study and 
\cite{gustafsson1999} are small for all these parameters. 
It should be noted that for these stars our
inclusion of the Fe\,{\sc i} blend barely has any effect on 
$\rm \Delta [C/H]$. The differences we see in $T_{\rm eff}$ (+12\,K) and 
$\log g$ (0.00\,dex) are small and would only translate into
$\rm \Delta [C/H]\approx 0.00$ and $\rm \Delta [Fe/H]\approx 0.01$ (see Table~\ref{tab:errors}).

In Fig.~\ref{fig:carbon_comparison} (right hand panel) we also plot  the
differences between our abundances and those from other works for a
total of 31 stars.  Comparisons are made to the two other studies that
use the  forbidden [C\,{\sc i}] line \citep{andersson1994, takeda2005}
as well as to studies that use the permitted C\,{\sc i} lines
\citep{tomkin1995, ecuvillon2004, takeda2005} and the molecular CH
bands \citep{laird1985}.  With \cite{andersson1994} we have 12 stars
in common, and for 4 of those they only give upper limit on
[C/H]. For the other 8 stars the difference is  $+0.12\pm0.12$\,dex,
i.e., quite large and substantially larger than relative to
\cite{gustafsson1999}. \cite{takeda2005} use both the forbidden and
the permitted carbon lines and we have 11 stars in common. The
mean difference between our and their forbidden [C/H] abundances is $+0.04\pm0.14$\,dex
and to their permitted [C/H] abundances it is
$-0.03\pm0.07$\,dex. With \cite{tomkin1995} we have 9 stars in common
and the difference is $+0.05\pm0.18$\,dex. With \cite{ecuvillon2004}
we have 13 stars in common and the difference is
$-0.08\pm0.08$\,dex. With \cite{laird1985} we have 11 stars in common
and the difference is $-0.08\pm0.19$\,dex

We would suggest that if one were to merge our sample with
the \cite{gustafsson1999} sample one should increase their [Fe/H] by
0.03\,dex and decrease their [C/H] by 0.02\,dex (or instead adjust our
values accordingly). Even though the reason(s) for these shifts are
not fully resolved this would at least bring our and their samples
onto a common  baseline. Merging our data with the \cite{andersson1994}
or the \cite{takeda2005} samples would introduce a lot of
extra scatter and make any interpretation of the observed carbon  trends
difficult.

Random errors are partly represented by uncertainties in the stellar
atmosphere parameters. In Table~\ref{tab:errors} we list how typical
uncertainties in effective temperature, metallicity, and surface
gravity ($\Delta T_{\rm eff}=+70$\,K, $\rm \Delta [Fe/H]=+0.1$, and
$\Delta \log g=+0.1$, respectively) effects the derived abundances for
three stars \citep[see also discussions in][]{bensby2003, bensby2004}.

Random errors may also be due to uncertainties, or defects, in the
observed spectra. Here we paid special attention to features in the
spectra that could be left-overs from the reduction procedure and a
possible insufficient removal of the fringing pattern. For seven stars
(listed  in Sect.~\ref{sec:carbonabundances}) we did not determine any
carbon abundances due to that the [C\,{\sc i}]  line was severely
deformed. For the remaining stars we could not distinguish any
possible fringing residuals from the ever-present random scatter
(noise). If the initial fringing pattern effects the final carbon
abundances (by an inefficient removal) we would probably see a
substantially larger scatter in the carbon trends than what we see in
our oxygen trends for which we used the same spectrograph and the same
analysis method but for the forbidden [O\,{\sc i}] line at
630.0\,nm \citep{bensby2004}. 
Although a larger scatter indeed  seem to be present for
the carbon trends than for our oxygen trends we find  that the [C/H]
vs [Fe/H] trend is quite well-defined (see Fig.~\ref{fig:ch_feh})  and
that effects due to residuals from the fringing pattern have been
reduced to a level where they most likely do not effect the general 
appearance of the abundance trends.

Random errors may also arise from errors in the actual fitting of the
synthetic  spectra to the observed ones. A major concern here is the
location of the continuum. Due to the weakness of the  [C\,{\sc i}]
line a small change in the location of the continuum will have an
effect on the carbon abundance.  This problem will vary with the
quality of the spectra as well  as the strengths of the [C\,{\sc i}]
line and surrounding lines making a true estimate of the uncertainties
in the placement of the continuum level difficult.  By examining our
spectra and varying the continuum level we estimate an average
uncertainty of 0.06\,dex in [C/H]. Since the spectra we used in
\cite{bensby2004} did not suffer from fringing effects and generally
were of higher quality we estimate the error in [O/H] to be half of
that in [C/H], i.e. 0.03\,dex.

Including a random error of 0.06\,dex in [C/H] and 0.03\,dex in [O/H]
due to uncertainties in the continuum level we estimate that the total
random errors in our final [C/H], [C/Fe], [O/H], and [C/O] are 0.07,
0.10, 0.07, and 0.07\,dex, respectively (see Table~\ref{tab:errors}).

\begin{figure*}
\resizebox{\hsize}{!}{\includegraphics[bb=18 144 592 555,clip]{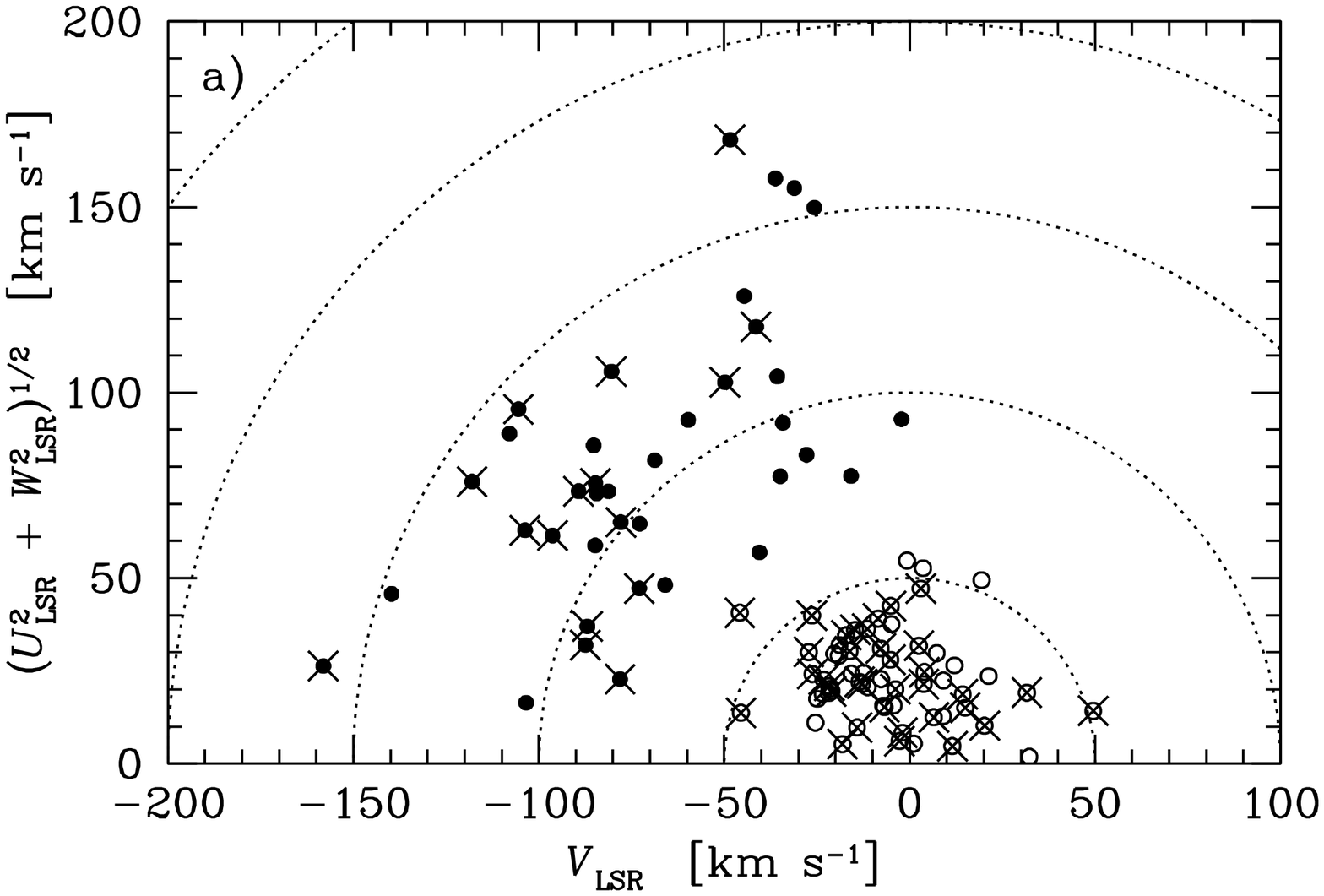}
                      \includegraphics[bb=18 144 592 555,clip]{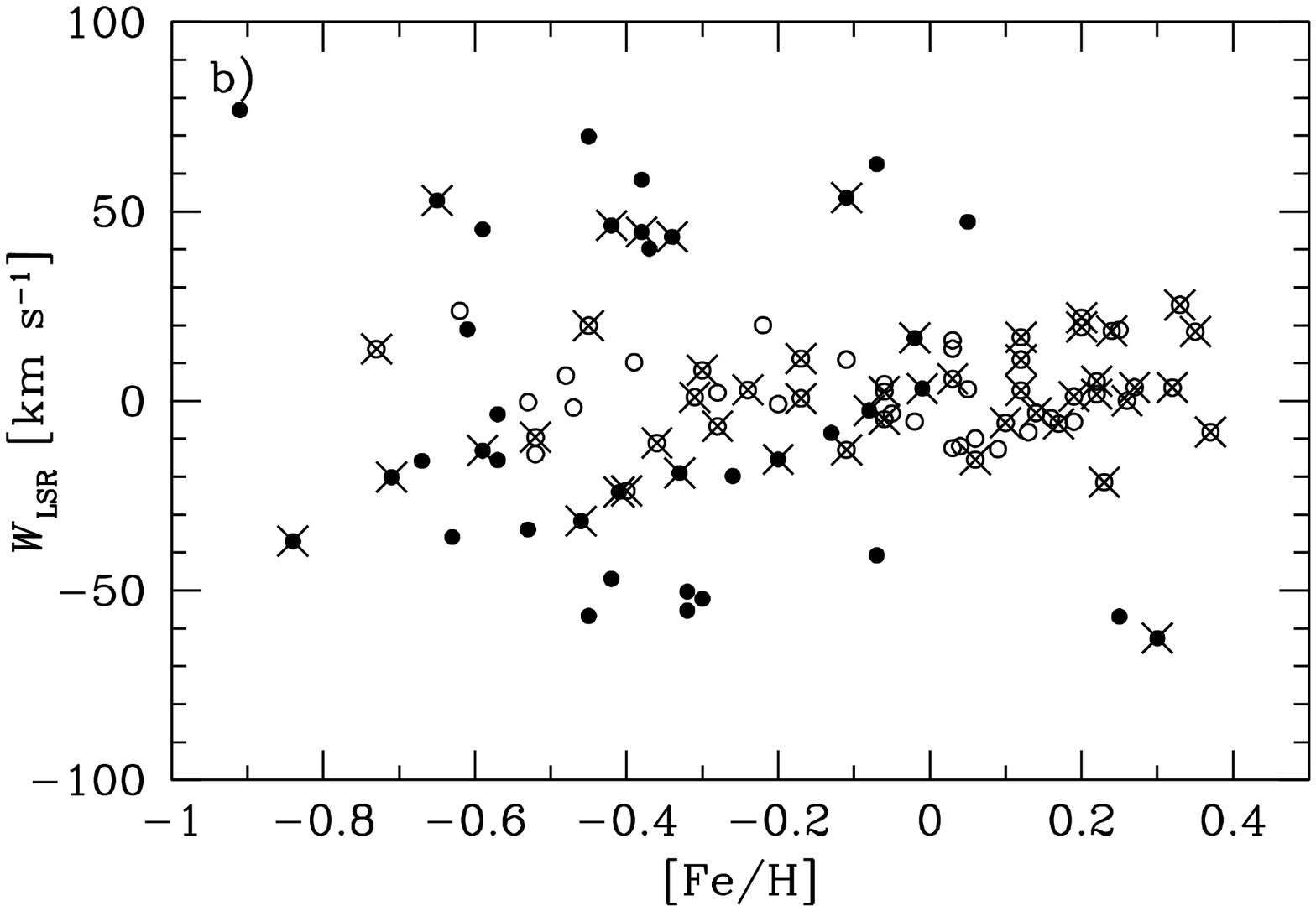}}
\caption{{\bf a)} Toomre diagram; dotted lines  indicate constant peculiar
         space velocities, $v_{\rm pec} = (U^{2}_{\rm LSR} +
         V^{2}_{\rm LSR} + W^{2}_{\rm LSR})^{1/2}$,  in steps of 50
         km~s$^{-1}$.
         {\bf b)} $W_{\rm LSR}$ versus [Fe/H].
         The plots include our full thin disk
         (empty circles) and thick disk (filled circles) samples from
         Bensby et al.~(2003, 2005). Those stars that are included in
         this study are marked by crosses.}
\label{fig:kinematics}
\end{figure*}

\section[]{Results and discussion} \label{sec:discussion}

Before we discuss our carbon results and where carbon might 
be produced we first, briefly,
recap some of the most important features of our sample selection, 
i.e., the kinematic definition of the two samples. We then turn to
a description of the abundance trends found and their implications
for where carbon can have formed.

\subsection{The kinematic definition of our samples} \label{sec:kinematic}

\begin{table}
\centering
\caption{Kinematic properties of the stellar populations that were used when
   calculating probabilities to select thin and thick disk stars.
The properties of the Hercules stream have been
                  adopted from Famaey et al.~(2005) and the properties of
                  the other populations are the same as in
                  Bensby et al.~(2005) except for the thin disk that has
                  a lower number density ($X$) due to the now included
                  Hercules stream.
All velocities and velocity dispersions are given in km\,s$^{-1}$.}
\label{tab:kinematics}
\centering
\setlength{\tabcolsep}{2mm}
\begin{tabular}{lrrrrrrr}
\hline
 & $X$    &
$\sigma_{\rm U}$ &
$\sigma_{\rm V}$ &
$\sigma_{\rm W}$ &
$U_{\rm lag}$ &
$V_{\rm lag}$ &
$W_{\rm lag}$ \\
\hline
Thin disk      & 0.84   &  35 & 20 & 16 &    0  &  $-15$ &   0   \\
Thick disk     & 0.10   &  67 & 38 & 35 &    0  &  $-46$ &   0   \\
Hercules       & 0.06   &  26 &  9 & 17 & $-40$ &  $-50$ & $-7$  \\
Halo           & 0.015  & 160 & 90 & 90 &    0  & $-220$ &   0   \\
\hline
\end{tabular}
\end{table}

The stars in our sample were selected, based on their kinematics, to
either be typical representatives of the thin or the thick Galactic
disk.  The kinematic selection is further described  in
\cite{bensby2003,bensby2005} including an extended discussion of how e.g. the
local normalisation of the thick disk influences the categorisation of
an individual star.

In Fig.~\ref{fig:kinematics}a we show a so called Toomre diagram for our
stars. The distribution of the stars in our two samples in this plot
is essentially  the same as \cite{fuhrmann2004} finds for his thin and
thick disk samples.  Figure~\ref{fig:kinematics}b shows the $W_{\rm LSR}$
velocity (which is proportional to the maximum distance below/above
the Galactic plane, $Z_{\rm max}$, a star can reach) as a function of
[Fe/H]. We see that even at $\rm [Fe/H]=0$ or higher there are  stars with
high $W_{\rm LSR}$. The number of such stars appear to  be higher than
one would expect from orbital heating by e.g. molecular  clouds
\citep{hanninen2002}. One should though remember that our sample is
far from complete and we certainly are subject to various biases.
However, our point  is that these stars do exist, are readily found in
any large catalogue (hence not rare objects)  and they imply that the
thick disk indeed extends to at least solar metallicities. 

Finally, it is also worthwhile to note, as pointed out by e.g.
\cite{nordstrom2004}, that there is a lot of kinematic structure when
studying large samples of nearby stars. One such structure, or group
of stars, of particular interest here is the Hercules stream which has 
kinematic properties very
similar to some of the thick disk stars. \cite{famaey2005} studied a
sample of  $\sim 6700$ nearby K and M giants and were able
to identify several such structures (including the Hercules stream) 
for which they determined
kinematic properties as well as local number densities. From their
results we see that it is especially Hercules stream stars with high
$U_{\rm LSR}$ velocities in the direction  away from the Galactic
centre that could  be erroneously classified as thick disk
stars. Using a compilation of chemical abundances \cite{soubiran2005}
find that the abundance trends for the Hercules stream mostly follow
those in the thin disk. 

We have checked
for possible contamination from Hercules stream  in our samples.
This was done by recalculating the probabilities that we used when 
selecting our
samples  \citep[see][]{bensby2003, bensby2005}, and now including the
properties for the Hercules stream (see Table~\ref{tab:kinematics}).
Using the same equations as in \cite{bensby2003, bensby2005} 
(but now  modified to also include 
$U_{\rm lag}$ and $W_{\rm lag}$) 
we find that none of our stars are more
likely to belong to the Hercules stream than to the thin or thick  disks.

For the discussion of our carbon results from this paper we will
assume that our stellar samples are representative for the thin and
thick disks, or the stellar components that can be kinematically
associated with the two disks disks, and that the thick disk does
extend up and until solar metallicities.

\subsection{Observed trends} \label{sec:trends}

\begin{figure*}
\resizebox{\hsize}{!}{\includegraphics[bb=18 144 592 485,clip]{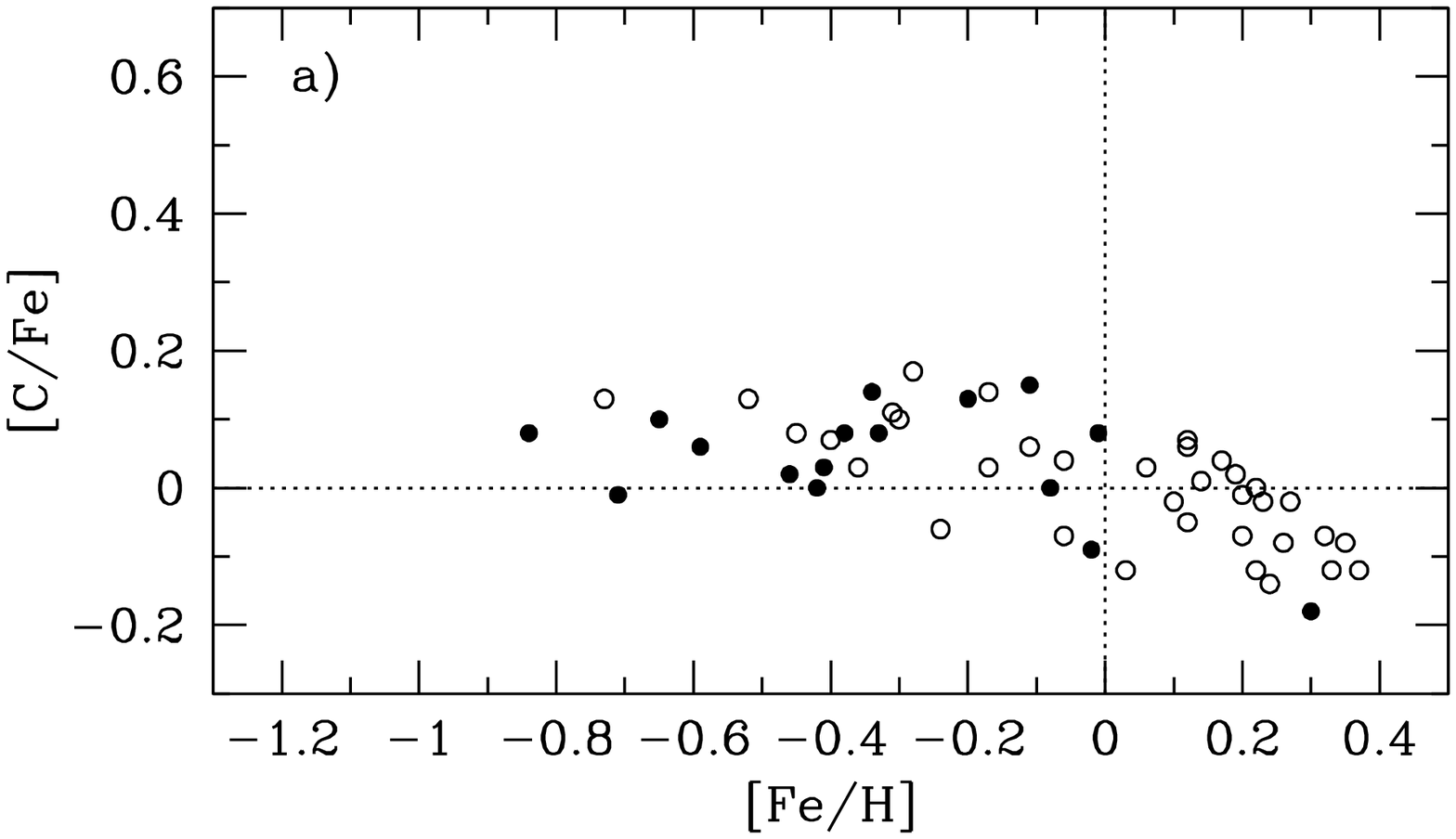}
                      \includegraphics[bb=18 144 592 485,clip]{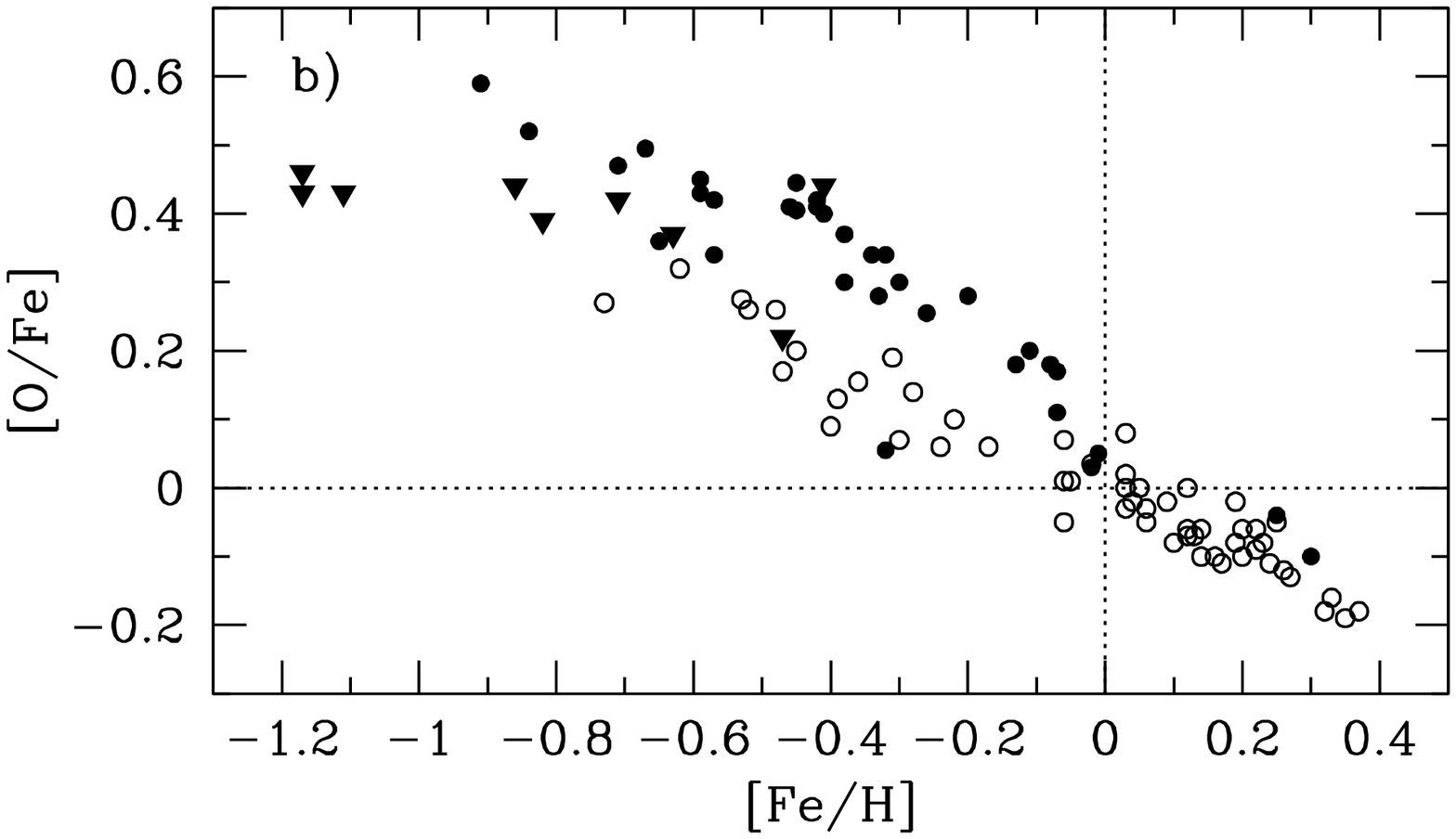}}
\resizebox{\hsize}{!}{\includegraphics[bb=18 144 592 530,clip]{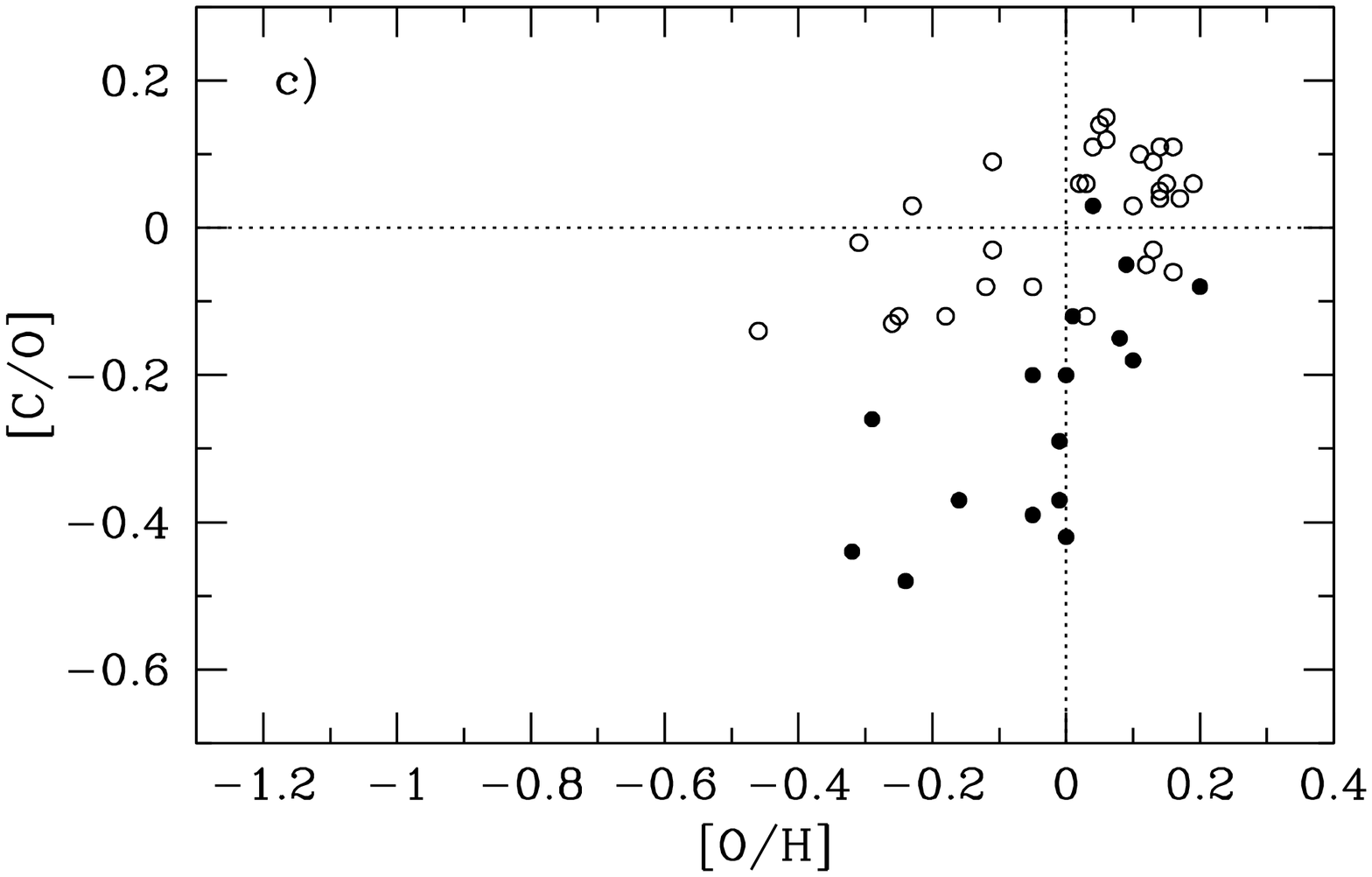}
                      \includegraphics[bb=18 144 592 530,clip]{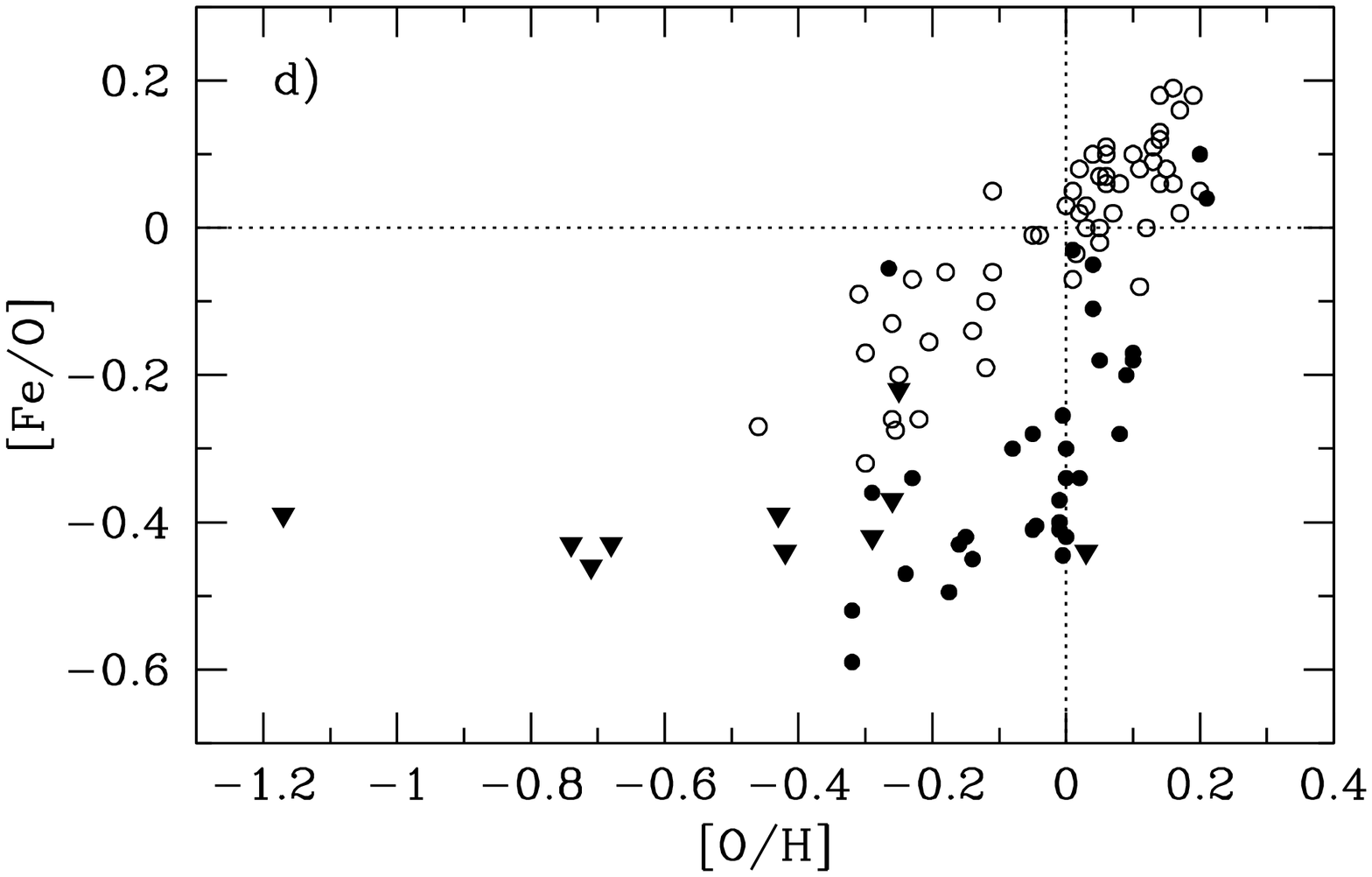}}
\caption{ {\bf a)} and {\bf b)} show the carbon and oxygen abundances,
        respectively, relative to iron. 
        {\bf c)} and {\bf d)} show the carbon and iron abundances,
        respectively, relative to oxygen. All carbon abundances are
        from this work and the oxygen and iron abundances have been
        taken from Bensby et al.~(2003, 2004, 2005), see also
        Table~\ref{tab:sampleA}.  Thin and thick disk stars 
        are marked
        by open and filled circles, respectively.  The few thick disk
        stars from Nissen et al.~(2002) included in the oxygen
        plots are marked by filled triangles. 
 }
\label{fig:trender}
\end{figure*}

\subsubsection[]{[C/Fe] versus [Fe/H]}

In Fig.~\ref{fig:trender}a we show the trend of [C/Fe] versus
[Fe/H]. The thin and thick disk [C/Fe] trends are fully merged and at
sub-solar metallicities the [C/Fe] values all fall within $\rm 0
\lesssim [C/Fe] \lesssim 0.2$ with no particular slope. At super-solar
metallicities the [C/Fe] values decrease with increasing [Fe/H]
(independent of if the Fe\,{\sc i} blend is included or not, compare
Figs.~\ref{fig:chdiff_noblend}d and e).

The flat [C/Fe] trend that we see at $\rm [Fe/H]<0$ is what generally
is found, within the uncertainties, in the literature  \citep[see,
e.g., compilation by][]{gavilan2005}.  \cite{gustafsson1999}, however,
find an increase in [C/Fe] for decreasing [Fe/H]. This is somewhat
discomforting since both our and their carbon abundances are based
on the forbidden line at 872.7\,nm.  However, at a closer look at the
[C/Fe] versus [Fe/H] trend in  \cite{gustafsson1999} it is possible
that ours and theirs trends are not that different. That they see a
slope seem to be mainly based on a few ($\sim 5$) stars with $\rm
[C/Fe]>0.2$ at $\rm [Fe/H]\lesssim -0.6$ (see their Fig.~4).
So, {\it if} those stars are disregarded, our  and their [C/Fe]
versus [Fe/H] trends are, within the uncertainties,  similar.
Good agreement to \cite{gustafsson1999} is further strengthened from 
the detailed comparison in  Sect.~\ref{sec:systematicerrors}.

\subsubsection[]{[C/O] versus [O/H]}

Fig.~\ref{fig:trender}c shows the [C/O] versus [O/H] trend for our
stellar sample. The thin and thick disks show trends
that are clearly separated.   The thin disk shows a shallow increase
in [C/O] with [O/H]  whilst the thick disk first has a flat [C/O]
trend that increases sharply at  $\rm [O/H]=0$. The great resemblance
with the observed [Fe/O] versus [O/H] trends shown in
Fig.~\ref{fig:trender}d could indicate that C and Fe indeed originate
from objects that evolve on similar time scales. As our stellar sample
spans a rather limited range in [O/H] it is difficult to use this
sample only to say how the [C/O] trend for the thick disk would have
evolved from very low oxygen abundances (i.e. $\rm [O/H]\lesssim-0.4$).  In
Fig.~\ref{fig:akerman} we therefore show our [C/O] versus [O/H] trend
together with the stellar sample (consisting of stars
with halo kinematics)  from \cite{akerman2004}. 
They based their carbon
and oxygen abundances on permitted C\,{\sc i} and O\,{\sc i} lines.
As they assumed that the NLTE corrections for the C and O
abundances for these lines are of similar size they did not apply
any NLTE corrections. However,
recent NLTE  calculations by \cite{fabbian2005} have shown that these
``permitted''  [C/H] values should be decreased by as much as 
$\sim 0.4$\,dex  at  $\rm [O/H]\approx-2.3$ and less at
 higher  [O/H]. The NLTE corrections for [O/H] are slightly
less ($-0.1$ to  $-0.3$\,dex, depending on stellar parameters) 
resulting in the [C/O] being overestimated by $\sim 0.2$\,dex at the lowest
[O/H] in the \cite{akerman2004} data \citep[see also][]{asplund2005araa}. 
As a consequence the upturn seen in [C/O] with decreasing [O/H]
for $\rm [O/H]\lesssim-1$ in the \cite{akerman2004} data
is overestimated - their trend should be flatter.   

\subsection{Where is carbon made?}

The carbon yields for various objects differ significantly.  The
yields are sensitive to a number of factors such as: stellar winds,
treatment of convection, and the $^{12}$C($\alpha$,$\gamma$)$^{16}$O
rate.  Earlier  models of massive stars found them to give high yields
of C \citep[e.g.,][]{maeder1992} whilst these were adjusted downwards
once rotation was taken into account \citep{meynet2002}. Several
studies have used these yields to model galactic chemical
evolution and compare it with observed  abundance trends (both in local
stars as well as in Galactic and  extra-galactic H\,{\sc ii}
regions). \cite{gustafsson1999} gave an exhaustive list and discussion
of  possible sites for carbon. These include: supernovae, novae,
Wolf-Rayet stars, low and intermediate mass stars in the planetary
nebula phase or by super-winds at the end or the red-giant
phase. Through an  analysis of Galactic dwarf stars  (a mixture of
thin disk stars, metal-poor thick disk stars, as well as stars with
intermediate kinematics from Edvardsson et al. 1993)  and data from
dwarf galaxies they arrived at the conclusion that the main
contribution of carbon to the galactic chemical evolution comes from
super-winds of metal-rich massive stars, that low mass stars were ruled
out, and that the  r\^ole of the intermediate mass stars remained
unclear. \cite{henry2000} modelled a large set of Galactic and
extra-galactic H\,{\sc ii} regions  and reached a similar conclusion.
Other studies find that the low  and intermediate mass stars are the
essential contributors to the  chemical enrichment of the Galaxy.
By adopting the new yields by \cite{meynet2002},
\cite{chiappini2003b} conclude that massive stars do contribute to the
carbon enrichment but that the main contribution at higher metallicities
must be due to low and intermediate mass stars.  Using the same set of
yields \cite{chiappini2003} are also able to explain the abundance ratios
in other galaxies, such as irregulars. 
The recent study by
\cite{carigi2005} arrives at a  similar complex explanation for the
observed abundance patterns where  metallicity dependent yields are
necessary both for the massive stars (increasing yields with
metallicity) as well as for low and intermediate mass stars
(decreasing yields with  increasing metallicity). In their
best-fitting model  the massive stars dominate the  C enrichment at
early times and the low mass and intermediate mass starting to
contribute later. The contributions at later times are roughly equal
for the two sources.  
As \cite{carigi2005} had to invoke the 
yields by \cite{maeder1992} 
\citep[that no longer should be used as they are superseded by][]{meynet2002} 
in order to reproduce the
solar [C/O] ratio the importance they attribute to massive stars at
higher metallicities could be overestimated. 

\begin{figure*}
\resizebox{0.8\hsize}{!}{
           \includegraphics[bb=40 170 610 385,clip]{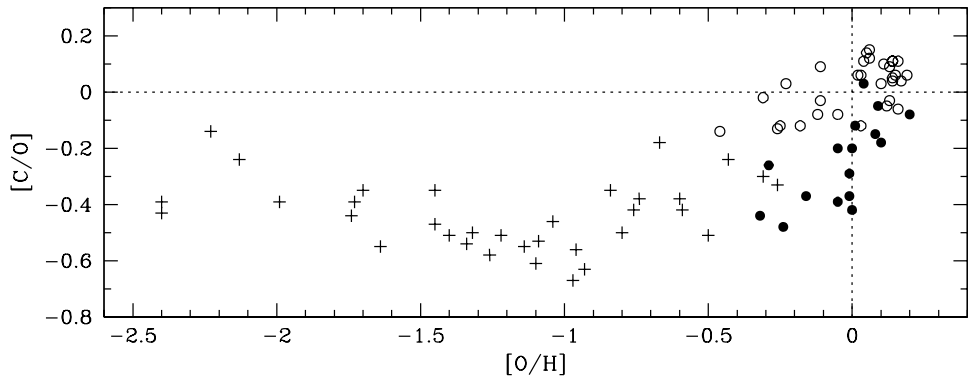}}
\caption{[C/O] versus [O/H]. Our thin and thick disk stars are marked
        by open and filled circles, respectively. Halo stars from
        Akerman et al.~(2004) are marked by crosses. Note that 
	the Akerman et al.~(2004) [C/O] values are
	overestimated by $\sim 0.2$\,dex at the lowest [O/H] 
	due to neglect of NLTE effects. The effect decreases as [O/H]
	increases.
        }
\label{fig:akerman}
\end{figure*}

Our [C/Fe] vs [Fe/H] trend in the thick disk is flat whilst  our
[O/Fe] vs [Fe/H] trend (see Figs.~\ref{fig:trender}a and b,
respectively)  shows a clear break and subsequent downturn at $\rm
[Fe/H] = -0.5$ (this corresponds to $\rm [O/H]\approx0$).  This type
of  downturn is normally interpreted as the onset of SN\,Ia
\citep[see, e.g.,][]{tinsley1979, matteucci2001}  and since oxygen is
only produced in SN\,II the increased Fe production by SN\,Ia results
in a downward trend as [Fe/H] increases.  For [C/Fe] vs [Fe/H] we see
no such trend. Hence, we may infer from the C abundances in our thick
disk sample that enrichment from carbon happens on the same time scale
as the enrichment from SN\,Ia. Hence the increase in Fe production is
matched by enrichment of C.

The fact that our [C/O]\,-\,[O/H] and [Fe/O]\,-\,[O/H] trends are
indeed very similar further strengthens the importance of low and intermediate
mass stars as contributors to the carbon enrichment at higher metallicities.
This is also supported by \cite{chiappini2003b} who
predict, if low and 
intermediate mass stars are important, that there first would be a flat
[C/O] plateau for the thick disk that then (at roughly solar [O/H]) should
sharply increase, to be followed by a shallow thin disk [C/O] trend at higher 
[C/O] ratios. And this is just what our data show.

Further, \cite{akerman2004} note that the amplitude of the rise that they see
in the [C/O] ratio around $\rm [O/H]=-1$ in their Milky Way halo
stars (see Fig.~\ref{fig:akerman}) is well matched by similar trends for 
H\,{\sc ii} regions in
nearby spiral and irregular galaxies. This they interpret as favourable
for their findings that massive stars are the main contributor to the
chemical enrichment of carbon as it would be very unlikely that the
abundance trends in galaxies that have experienced different types of
chemical histories should be so similar if carbon was mainly produced
in sites where a substantial time-delay had to be invoked, e.g. low-mass 
stars. Given the uncertainties relating to NLTE corrections of the 
\cite{akerman2004} data (see discussion above)
and that \cite{chiappini2003} are able to explain
the abundance ratios in other galaxies as well (including irregulars),
using the same set of yields that were used for the Milky Way, 
this indicates that the r\^ole of massive stars might be overestimated by
\cite{akerman2004}.

If we compare our C abundances with Y the  trend of [C/Y] versus [Y/H]
shows a flat trend for $\rm [Y/H]\lesssim 0$ which turns to a steadily
declining trend for $\rm [Y/H]\gtrsim 0$ (see Fig.~\ref{fig:cyfeo}a)
which is very similar to what we see for [C/Fe] versus [Fe/H] (see
Fig.~\ref{fig:trender}a).  [C/Y] vs [Fe/H] is somewhat different,
being, essentially,  flat for both the thin and the thick disk samples
at all metallicities (see Fig.~\ref{fig:cyfeo}b).  Approximately 70\%
of Y is produced in the s-process in AGB stars and  $\sim 30$\% comes
from massive stars ($\sim 10$\% from a weak ``secondary''  s-process
component and $\sim 20$\% from a primary component
\citealt{travaglio2004}).  The most straightforward interpretation of the
lack of trends would  be that the two elements, C and Y, are made in
objects that enrich the interstellar medium on the same time
scale. Most naively we would  assume that their major components  are
indeed made in the same objects, namely low and intermediate mass
stars in the AGB phase.  However, if we inspect the  trend of [C/Y]
versus [Y/H] in addition to the flat trend below sub-solar  [Y/H] we
see a declining trend of [C/Y] for [Y/H]$>0$ indicating that as the Y
increases there is a decrease in C production. This could be due to a
metallicity effect in the production of s-process elements that
favours light s-process elements (such as Y) over heavy s-process
elements (such as Ba) at high metallicities
\citep{busso2001,bensby2005}.

So, we have shown that the carbon production must balance the
the Fe production by SN\,Ia and the Y production by AGB stars. 
With this observable in mind it appears
that either the massive stars must have (very) metallicity
dependent yields that are finely tuned to the [Fe/H] timescale of
SN\,Ia as observed in the Milky Way thick disk or that less massive 
stars are significant contributors too. 

In summary we thus find that 
the carbon enrichment at low metallicities (i.e. the Galactic halo and 
metal-poor thick disk) is due to massive stars but as 
more and more low and intermediate mass stars start to contribute to the
general enrichment they also more and more dominate the
carbon enrichment of the interstellar medium as it is seen in the
thin disk and the metal-rich thick disk.  

\begin{figure}
\resizebox{\hsize}{!}{\includegraphics{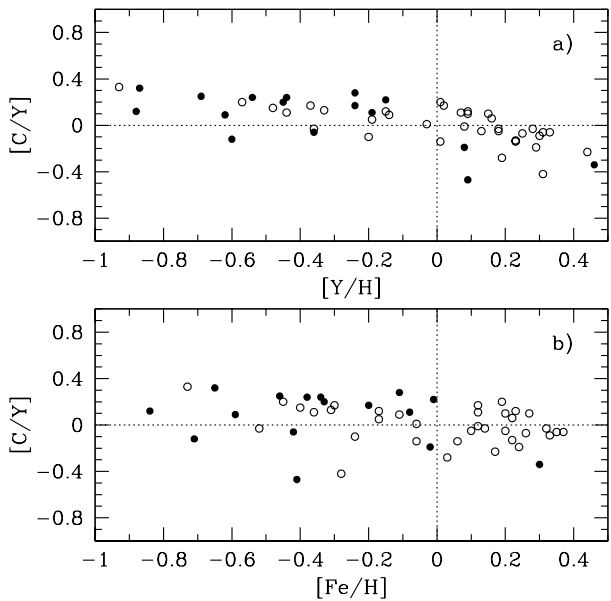}}
\caption{
        {\bf a)} [C/Y] versus [Y/H] and {\bf b)} [C/Y] versus [Fe/H].
        Yttrium abundances are taken from Bensby et al.~(2005).
        Thin and thick disk stars are marked by open and filled circles,
        respectively.
        }
\label{fig:cyfeo}
\end{figure}

\section{Summary} \label{sec:summary}

We present carbon abundances in 51 F and G dwarf stars (16 thick disk stars
and 35 thin disk stars)in the solar 
neighbourhood. The analysis is based on the forbidden [C\,{\sc i}] line at 
872.7\,nm which is an abundance indicator that is insensitive to errors
in the stellar atmosphere parameters. Combining these data with our previously
published oxygen abundances, based on the forbidden [O\,{\sc i}] line at
630.0\,nm \citep[see][]{bensby2004}, we can form very robust [C/O] ratios 
that we then used to investigate the origin of carbon and the chemical 
evolution of the Galactic thin and thick disks. 
Our findings concerning carbon abundance trends in 
the Galactic thin and thick disks include:
\smallskip\\\noindent$\bigstar$ 
At sub-solar [Fe/H] we find that the [C/Fe] versus [Fe/H] trends
for the thin and thick disks are fully merged and essentially flat.
For the thin disk, that extends to higher metallicities, [C/Fe] is flat until 
$\rm [Fe/H]=0$ when a slow decline starts (see Fig.~\ref{fig:trender}a).
\smallskip\\\noindent$\bigstar$
Our abundance trends indicate that the sources that contribute to the
carbon enrichment of the interstellar medium do so on the same time-scale
as those that produce most of the iron, i.e. SN\,Ia. The production of 
C and Y (coming mainly from AGB stars) also seem to work on the same 
time-scale.
\smallskip\\\noindent$\bigstar$
In light of our own as well
as other studies in the literature 
\citep[notably][]{gustafsson1999, chiappini2003, chiappini2003b, 
akerman2004, carigi2005} we feel that the source(s) of carbon is not
yet settled but that there is growing evidence that a complicated,
and finely tuned, set of objects contribute to the enrichment of carbon
in galaxies.
As discussed, based on our own results only we would conclude 
that the main source for carbon in the Galaxy is low and intermediate 
mass stars.
However, for the Milky Way galaxy it appears that massive stars played a 
significant r\^ole for the carbon enrichment at low metallicities 
(i.e. halo and metal-poor
thick disk) whereas low and intermediate mass stars dominate more and more 
at higher metallicities, i.e. that they have been the major contributors to the 
carbon enrichment in the thin disk and the metal-rich thick disk.
\smallskip\\\indent
From our analysis of the
forbidden [C\,{\sc i}] line at 872.7\,nm to determine carbon abundances in
solar-type stars we also found a few details that we think are noteworthy:
{\bf 1)}
By examining how the neglect of the blending Fe\,{\sc i} line
effects the derived carbon abundances from the forbidden [C\,{\sc i}] line
we find that it is mainly for stars with effective temperatures less than
$\sim 5700$\,K that differences are large
(see Fig.~\ref{fig:chdiff_noblend}a);
{\bf 2)}
Currently, the only available source for the oscillator strength of the
blending Fe\,{\sc i} line is a theoretical one \citep{kurucz1993}.
Therefore, it is possible that the strength of the Fe\,{\sc i} could
under- or overestimated. By working strictly relative to the Sun the
effects of an erroneous
treatment of the blend will be reduced. {\it It would, however, be
very valuable if a $\log gf$-value for the Fe\,{\sc i} line could be measured
in the laboratory};
{\bf 3)}
For the Sun we find a carbon abundance of $\rm \epsilon_{\odot}(C)=8.41$
when including the Fe\,{\sc i} blend and $\rm \epsilon_{\odot}(C)=8.44$
when neglecting it;
{\bf 4)}
The general appearance of our carbon trends in the two disks is
similar whether we include the Fe\,{\sc i} blend
in the abundance analysis or not (see Figs.~\ref{fig:chdiff_noblend}d and e).

\section*{Acknowledgements}

Thomas Bensby acknowledges support from the National Science
Foundation, grant  AST-0448900. Sofia Feltzing is a Royal Swedish
Academy Research Fellow supported by a grant from the Knut and Alice
Wallenberg Foundation.  We would like to thank the developers of
the Uppsala {\sc marcs} code, Bengt Gustafsson, Kjell Eriksson, Martin
Asplund, and Bengt Edvardsson who we also thank for providing us with the
{\sc spectrum} line synthesis program.
We also thank our referee, Martin Asplund, for valuable comments that
improved the  paper. Bengt Edvardsson,
Christina Chiappini, and Jacco van Loon are also thanked for
reading and giving constructive response on draft versions of the paper.
This research has made use of the SIMBAD database,
operated at  CDS, Strasbourg, France.

\bibliographystyle{mn2e}
\bibliography{referenser}

\appendix

\bsp

\label{lastpage}

\end{document}